\documentclass[10pt,journal,compsoc]{IEEEtran}
\IEEEoverridecommandlockouts
\usepackage{cite}
\usepackage{amsmath,amssymb,amsfonts}
\usepackage{algorithmic}
\usepackage{graphicx}
\usepackage{textcomp}
\usepackage[linesnumbered,ruled,vlined]{algorithm2e}
\usepackage{multirow}
\usepackage[table,xcdraw]{xcolor}
\usepackage{subfigure}
\usepackage[numbers,sort&compress]{natbib}

\usepackage{hyperref} 

\newtheorem{Proposition}{Proposition}{}

{}
{}
{}
\newtheorem{proof}{Proof}

\usepackage{cite}
\def\BibTeX{{\rm B\kern-.05em{\sc i\kern-.025em b}\kern-.08em
    T\kern-.1667em\lower.7ex\hbox{E}\kern-.125emX}}

\begin{document}
\title{Cost Sensitive GNN-based Imbalanced Learning for Mobile Social Network Fraud Detection\\}
\author{Xinxin~Hu,
        Haotian~Chen,
        Hongchang~Chen,
        Shuxin~Liu,
        Xing~Li,
        Shibo~Zhang,
        Yahui~Wang,
        and Xiangyang~Xue, ~\IEEEmembership{Member,~IEEE}
\IEEEcompsocitemizethanks{\IEEEcompsocthanksitem Xinxin Hu, Hongchang Chen, Shuxin Liu, Xing Li, Shibo Zhang and Yahui Wang are with National Digital Switching System Engineering and Technological Research Center, Zhengzhou, China, 450002.(E-mail:  hxx@alumni.hust.edu.cn, ndscchc@139.com, liushuxin11@126.com, lixing\_ndsc@163.com, suzzyndsc@163.com, wangyahuindsc@163.com)\protect\\

\IEEEcompsocthanksitem Haotian Chen is with The Edward S. Rogers Sr. Department of Electrical \& Computer Engineering, University of Toronto, Toronto, Canada, M5S 3G4.(E-mail: hhaotian.chen@mail.utoronto.ca) \protect\\

\IEEEcompsocthanksitem Xiangyang Xue is with Institute of Big Data, Fudan University, Shanghai, China.(E-mail: xyxue@fudan.edu.cn)} \protect\\

\thanks{This work is supported by Henan Province Major Science and Technology Project under Grant 221100210100, Central Plains Talent Foundation of China under Grant 212101510002 and the National Natural Science Foundation of China under Grant 61803384.}
\thanks{Corresponding author: Hongchang Chen, Shuxin Liu}
\thanks{Xinxin Hu and Haotian Chen contributed equally to this work and should be considered co-first authors.}

\thanks{Manuscript received XX XX, XXXX; revised XX XX, XXXX.}}
\maketitle

\markboth{IEEE Transactions on Computational Social Systems,~Vol.~14, No.~8, Dec~2022}%
{Shell \MakeLowercase{\textit{et al.}}: Bare Demo of IEEEtran.cls for Computer Society Journals}

\begin{abstract}
With the rapid development of mobile networks, the people's social contacts have been considerably facilitated. However, the rise of mobile social network fraud upon those networks, has caused a great deal of distress, in case of depleting personal and social wealth, then potentially doing significant economic harm. To detect fraudulent users, call detail record (CDR) data, which portrays the social behavior of users in mobile networks, has been widely utilized. But the imbalance problem in the aforementioned data, which could severely hinder the effectiveness of fraud detectors based on graph neural networks(GNN), has hardly been addressed in previous work. In this paper, we are going to present a novel \underline{C}ost-\underline{S}ensitive \underline{G}raph \underline{N}eural \underline{N}etwork (CSGNN) by creatively combining cost-sensitive learning and graph neural networks. We conduct extensive experiments on two open-source real-world mobile network fraud datasets. The results show that CSGNN can effectively solve the graph imbalance problem and then achieve better detection performance than the state-of-the-art algorithms. We believe that our research can be applied to solve the graph imbalance problems in other fields . The CSGNN code and datasets are publicly available at \textcolor{blue}{\url{https://github.com/xxhu94/CSGNN}}.
\end{abstract}


\section{Introduction}\label{introduction}
\IEEEPARstart{M}{obile} social network fraud, also known as telecom fraud, is the misuse of mobile network products and services with the malicious intention  to extort money from victims. In August 2016, a professor in Beijing, China, was scammed out of $\$$267 million by a telecom fraudster disguised as a judicial official \cite{yang2019mining}. In addition to financial losses, a telecom fraud even claimed the life of a Chinese high school graduate in the same year \cite{zheng2018generative}. In addition, mobile social network fraud can spark public opinion as well. Cheong Wa Dae, the presidential office of South Korea, publicly responded to several cases of telecom fraud impersonating relatives of the president, in which people were even defrauded of hundreds of millions of won \cite{caict2020fraud}. Now, the prevalence of mobile social network fraud appears to be a worldwide phenomenon. Furthermore, fraud scripts have been evolving with social hotspots, leaving people more susceptible to the telecom fraud. According to Truecaller report in May 2022 \cite{truecaller2022spam}, up to 68.4 million Americans (26$\%$ of the American population ) lost a whopping 39.5 billion because of the telecom fraud, which is higher than the number (59.4 million) in 2021. One of three Americans (33$\%$) have fallen victim to a telecom fraud, and 20$\%$ of Americans have been scammed more than once. Based on the data provided by Whoscall in August 2022\cite{whoscall2022latest},  the total number of fraudulent phone calls and text messages across the world is around 460 million inquiries in 2021, a 58$\%$ increase over the previous year. From 2021 to 2022, the world has been experiencing an urgent economic recovery  under the shadow of COVID-19. In this  circumstances, disease prevention policies, vaccinations, relief programs, stock markets, and even virtual currencies have become  popular fraudulent themes. The surge in the number of scams has become a new normal in the post-pandemic world.

In the following arguments, we will explain why it is difficult to accurately detect mobile social network fraud  when a  plethora of data is available. First of all, the academic research on mobile network social fraud detection is generally unexplored due to the accessibility and high sensitivity of telecom data. The majority of current fraud detection methods \cite{yang2019mining, hooi2016fraudar,tseng2015fraudetector} conducted their experiments on synthetic or unpublished real-world data, making it challenging for other researchers to replicate the findings. Furthermore, since call detail records (CDR) usually are unstructured data, using a grid data processing approach to handle them can lead to the loss of a lot of structural information, which can greatly reduce the detection accuracy. Fortunately, the rise of graph machine learning techniques has provided new solutions for unstructured data processing in telecom fraud detection. However, the graph imbalance problem has deteriorated the performance of graph machine learning models in telecom fraud detection problems.

\begin{figure}[h]
  \includegraphics[width=\linewidth]{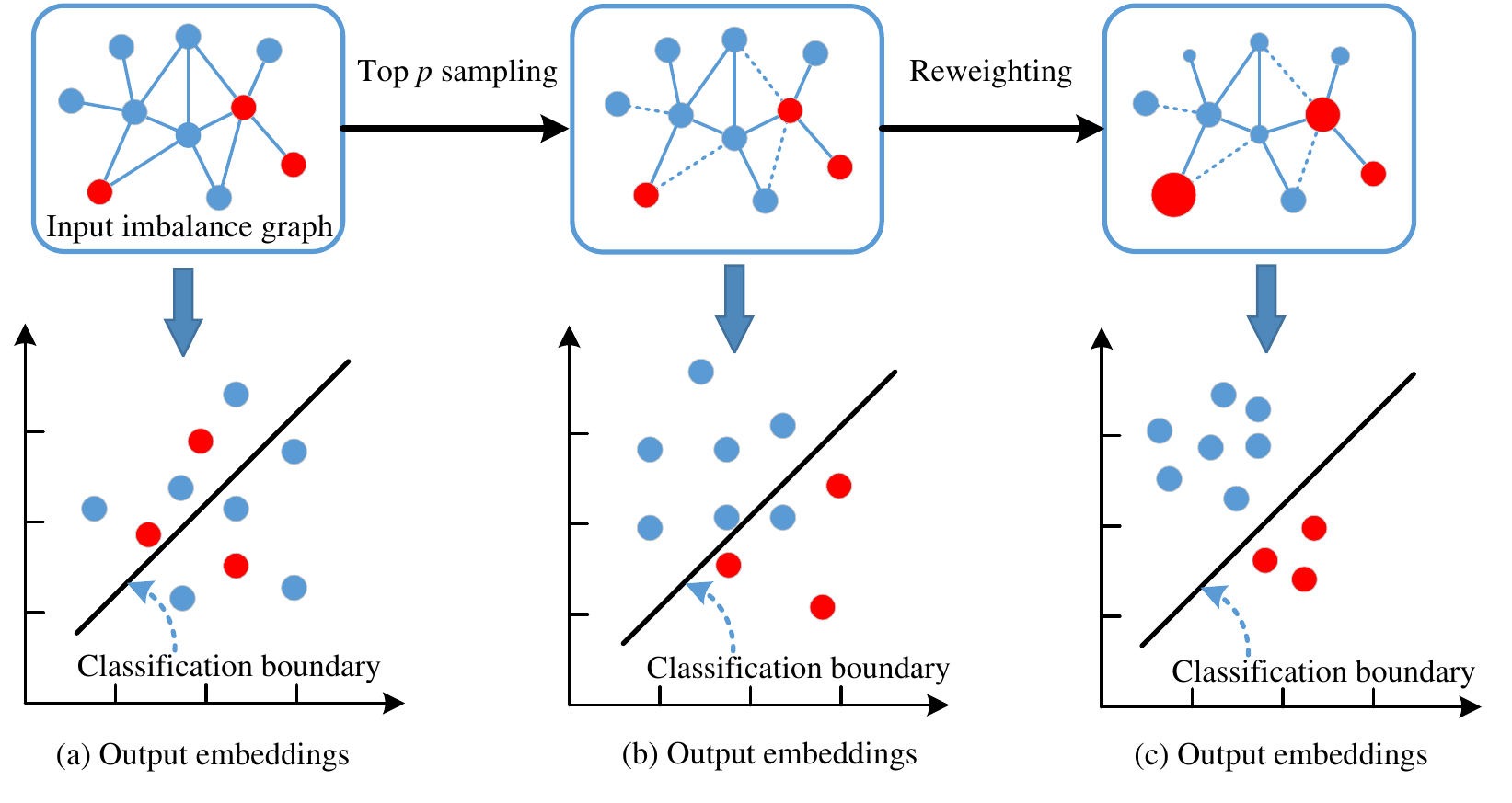}
  \caption{Illustration of the graph imbalance problem in fraud detection and the idea of CSGNN. Blue nodes indicate the benign and red indicates the fraud. a) The output embeddings of the input graph are difficult to distinguish benign from fraud. b) The situation improves after top-$p$ sampling-based aggregation of domain information. c) After cost-sensitive reweighting, the inter-class distance is larger and the intra-class distance is smaller.}
  \label{imbalance}
\end{figure}

Specifically, the problem of graph imbalance is that the number of positive samples and negative samples in the graph are unbalanced, or even severely biased. For example, fraudulent users in telecommunication networks only account for a small fraction of all users. Yang et al. \cite{yang2019mining} also pointed out that the data imbalance problem in telecom fraud detection has not been properly researched. In fact, the graph imbalance problem is not only occurred in the field of telecom fraud detection but also is widely existed in the fields of social fraud, insurance fraud, financial fraud, and cyber security. However, the graph imbalance, as a brandnew problem, has received little attention in current research. Though it is especially important in scenarios such as graph anomaly detection, likefraud detection. Usually, the graph imbalance problem is difficult to solve because there are two challenges: 1) the positive nodes have too many negative neighbors, thus affecting the model judgment during the aggregation of GNN neighbor information; 2) The proportion of positive nodes to all nodes is too low for the model to learn and then cannot effectively identify the corresponding patterns (see Figure \ref{imbalance} (a)). 

In order to solve the graph imbalance problem, most of the existing works only concern about challenge 2) and overcome it through oversampling or under sampling methods \cite{huang2022graph,wu2021graphmixup,zhao2021graphsmote,liu2021pick}. However, the oversampling methods may lead to overfitting, and the undersampling methods can cause the model to discard valuable information. Cost-sensitive learning is an effective method to handle the imbalance problem in the field of computer vision and other grid data \cite{khan2017cost}. Therefore, it seems to be a promising solution to graph imbalance problems. However, it can be challenging to employ the cost-sensitive learning on graph data efficiently.

To address the both challenges of the graph imbalance problem simultaneously, we propose a novel cost-sensitive graph neural network model CSGNN (see Figure \ref{imbalance} (b)(c)) that combines GNN with reinforcement learning(RL)-based neighbor sampling (for challenge 1)) and cost-sensitive learning (for challenge 2)). In particular, we first design a reinforcement learning strategy for node neighbor selection to balance the number of positive and negative neighbors of each node. Subsequently, we deliver the sampled graph to the GNN for embedding learning. Finally, the final learned embeddings are transferred to a well-designed cost-sensitive learner, which can learn cost-sensitive node embedding representations. In case of the cost sensitive learner, we propose a method for automated cost sensitive matrix learning, optimized by the back-propagation, and the theoretical proof will be provided. Extensive experiments on real-world public telecom fraud detection datasets demonstrate that the proposed scheme in this paper can achievecompetitive performance.

The main contributions of this paper can be summarized as follows:
\begin{itemize}
\item A novel framework is specifically designed for cost-sensitive graph neural networks that can efficiently solve the issue of the graph imbalance in mobile social network fraud detection;
\item In order to sample the imbalanced neighbors in graphs properly, a new reinforcement learning-based method is developed;
\item An automated cost matrix learning method based on gradient descent is proposed and the effectiveness of its combination with GNN is proved theoretically;
\item Extensive experiments on two real-world telecom fraud detection datasets demonstratethe superiority of the proposed method.
\end{itemize}

\section{RELATED WORK}
In this section, we briefly describe the related works, including imbalanced graph learning and mobile social network fraud detection.

\subsection{Imbalance Graph Learning}
Methodologically, our research belongs to imbalance graph learning, which is widely used in the fields of fraud detection, network security, anomaly detection, etc.  In recent years, more and more works have focused on the imbalance problem in graphs, and these works can be classified methodologically into the following categories:

The first category is based on graph reconstruction. Concretely, such methods reconstruct the graph by the generation of nodes and edges or the sampling of nodes and edges, and then perform representation learning on the reconstructed graph. GNNINCM \cite{huang2022graph} adopted embedding clustering-based optimization (ECO), graph reconstruction-based optimization (GRO), and hard sample-based distribution alignment losses for supervised training. Wu et al.\cite{wu2021graphmixup} constructed the semantic relational space that allows feature mixup at the semantic level, followed by synthetic minority node generation. Additionally, they used context-based self-supervision to capture local and global information in graph structures and then performed the edge Mixup. Wang et al. \cite{2021Network} proposed two imbalanced graph learning methods, RSDNE and RECT. The shallow method, RSDNE, guarantees intra-class similarity and inter-class dissimilarity in an approximate manner. In contrast, RECT explores class semantic knowledge, measuring structural loss on the reconstructed graph and semantic loss in the semantic space, which allows RECT to process networks with node features and multi-label settings. GraphSMOTE \cite{zhao2021graphsmote} combined GNN with the classical oversampling algorithm SMOTE to construct an embedding space to encode the similarity between nodes and synthesize new nodes in this space. Furthermore, the authors construct an edge generator to model the relational information for the newly generated nodes. PC-GNN \cite{liu2021pick} performed imbalanced graph learning through undersampling and oversampling of different classes of nodes and selective aggregation of neighbor information.

The second category is based on cost-sensitive learning. These methods combine GNN and cost-sensitive learning by giving greater weight to the minority class, thus allowing the model to better learn the features of the minority class. FACS-GCN \cite{santos2022facs} combined cost-sensitive exponential loss and adversarial learning components with a staged additive modeling approach to ensure the performance of GNN on unbalanced graphs. The dual cost-sensitive graph convolutional network (DCSGCN)\cite{duan2022dual} contained two sub-networks that compute the posterior probability and the cost of misclassification, respectively. It uses the cost as supplementary information in the classification to correct the posterior probability from the minimal risk perspective.

The third category is hybrid methods, which combine multiple methods for imbalanced graph. The graph classification model, G$^2$GNN \cite{wang2022imbalanced}, adopted techniques including similarity-based graph construction, oversampling minority classes, and undersampled graph enhancement based on nodes and edges to perform imbalance graph learning. The graph neural network framework based on curriculum learning (GNN-CL) was constructed by Li et al. \cite{li2022graph}. First, they obtained certain reliable interpolation nodes and edges by graph sampling based on smoothness and homogeneity. Then graph classification loss and metric learning loss were combined to adjust the distances between different nodes in the feature space associated with the minority category. mGNN\cite{liu2021pick} combined oversampling and cost-sensitive learning, performs feature enhancement, oversampling, and other operations on minority nodes, and gives more weight to minority nodes so that GNN can better adapt to the learning of imbalanced graphs.

Other methods: Wang et al. \cite{wang2021distance} proposed a class prototype-driven balanced training scheme (DPGNN) based on episodic training, supplemented with distance metric learning and imbalanced label propagation, to achieve imbalanced node classification in graphs. Shi et al.\cite{shi2020multi} proposed Dual Regularized Graph Convolutional Networks (DRGCN), which solves the imbalance representation learning of classes through the class conditioned adversarial training process and two kinds of regularization operation.

\subsection{Mobile Social Network Fraud Detection}
Empirically, our research pertains to mobile social network detection. In recent years, more and more work has been conducted on mobile social network fraud detection.

Akshaya et al.\cite{ravi2022wangiri} defined three Wangiri fraud patterns and assessed the safety and performance of unsupervised (One Class Support Vector Machine, Isolation Forest, Auto-encoders) and supervised (Naive Bayes, Random Forest, XGBoost) machine learning (ML) methods in detecting Wangiri patterns. In general, the supervised algorithms outperform the other algorithms, but for different Wangiri fraud modes, the optimal detection methods are not the same. Yang et al.\cite{yang2019mining} revealed the phenomenon of "precision fraud" in telecom fraud and the tactics fraudsters use to pinpoint targets, so as to determine that the user's information may have been seriously compromised, and then designed a semi-supervised detection framework, which contains Attribute factor, Macro interactive factor, Micro interactive factor, and Group factor. To detect telecom fraudsters from sequence behaviors, Jiang et al.\cite{jiang2022telecom} extracted source and target neighbor sequences from a sequential dichotomous network of user behaviors, and proposed a new Hawkes-enhanced sequence model (HESM) to integrate Hawkes-process into LSTM for historical influence learning. Targeting the communication sparsity of people's phone calls and text messages in CDR, Hu et al.\cite{hu2022btg} proposed a graph machine learning framework for sparse graphs to detect telecom fraud. After the extraction of user behavior features and pattern features, they constructed the graph based on similarity and then applied graph machine-learning techniques for fraud detection. To solve the imbalance problem in telecom fraud, Krasi{\'c} et al.\cite{krasic2022telecom} first adopted reinforcement learning for unbalanced neighbor sampling and then used GNN for information aggregation.

\begin{figure*}[h]
  \includegraphics[width=\linewidth]{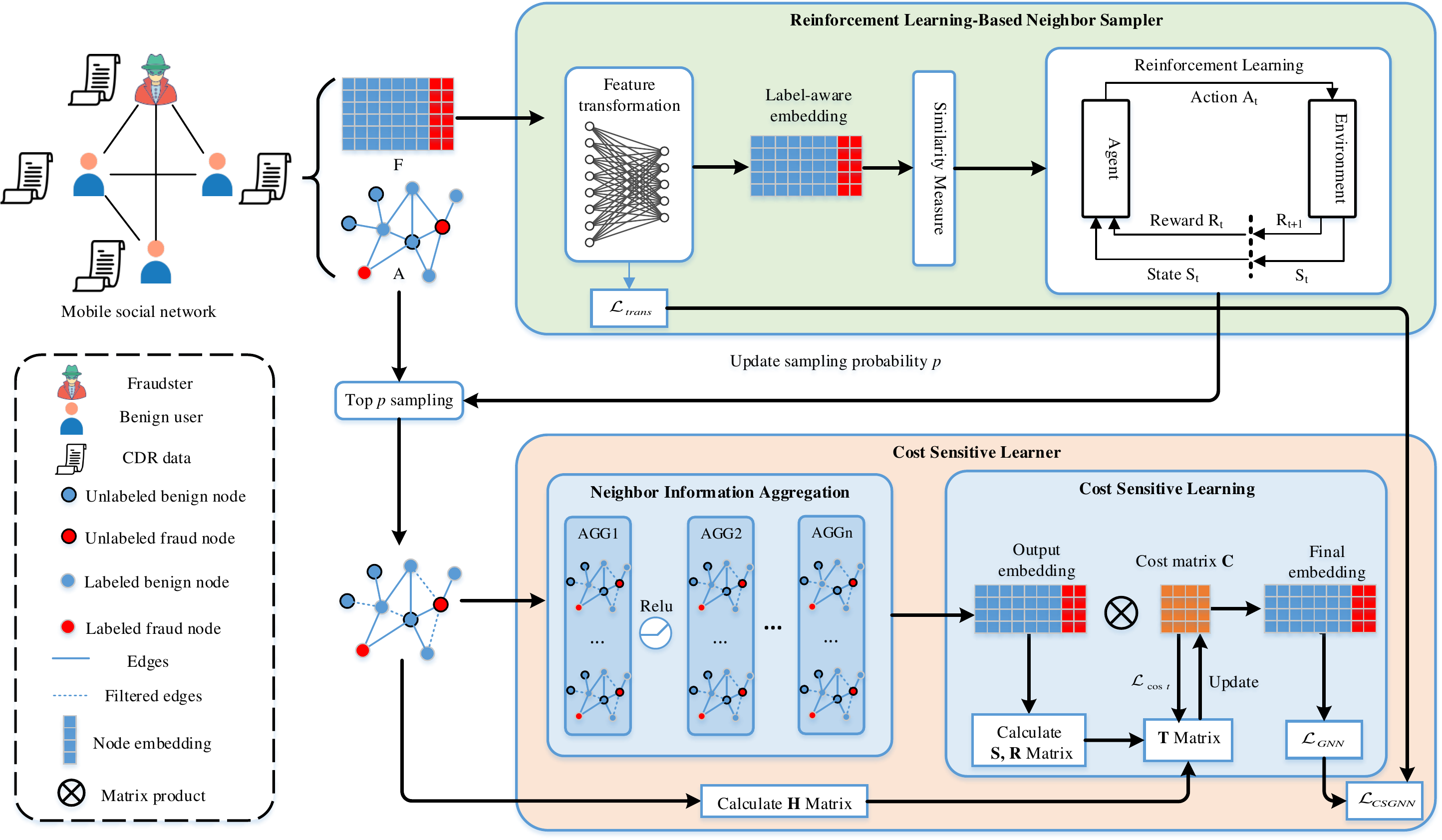}
  \caption{The overview of our proposed model CSGNN.}
  \label{CSGNN}
\end{figure*}

\section{Preliminary}\label{Preliminary}
In this section, we will introduce the definition of graph imbalance problem, cost-sensitive learning and GNN-based social network fraud detection. And all the essential notations are summarized in Table \ref{notation}.

\begin{table}[htbp]
\caption{Glossary of Notations.}
\centering
\renewcommand{\arraystretch}{1.5}
\setlength{\tabcolsep}{1.3mm}{
\begin{tabular}{|l|l|}
\hline
Symbol & Definition  \\ \hline
$\mathbf{A}$  & Adjacency matrix  \\ \hline
$\mathbf{X}^{(l)}$& Input feature matrix of the l-th layer,X$^{(0)}$=X  \\ \hline
$l, L$    & GNN layer number; Total number of layers    \\ \hline
$e, E $ & Training epoch number; Total number of epochs   \\ \hline
$\mathbf{W}$ & Neural Network Weight  \\ \hline
$\mathbf{h}^{(l)}$ & Node embedding at layer $l$  \\ \hline
$\mathbf{z}$& The final node embedding of GNN  \\ \hline
$\mathbf{C}$& Cost sensitive matrix \\ \hline
$\mathbf{C}_{i,j}$  & The cost of classifying node in class $i $ as class  $j $ \\ \hline
$\mathcal{N}$,$\mathcal{N}_i$& Number of samples; Number of samples in class $i$ \\ \hline
$K$ & Number of sample classes \\ \hline
$\mathcal{S}$ & The state space \\ \hline
$\mathcal{A}$ & The action space \\ \hline
\end{tabular}}
\label{notation}
\end{table}

\textbf{Definition 2.1. Graph Imbalance Problem.}
Given the label $\mathcal{Y}$ of a set of nodes in a graph, there are a total of $K$ classes,  $\left \{ C_{1}, . . ,C_{K} \right \}$.  $\left | C_{i}  \right |$  is the size of the $i$-th class, i.e., the number of samples belonging to that class. We use
 \begin{equation}
\mathrm{IR}=\frac{\min _{i}\left(\left|C_{i}\right|\right)}{\max _{i}\left(\left|C_{i}\right|\right)}
\end{equation}
to measure the class imbalance ratio. Therefore, $IR$ is located in the range [0, 1]. If $IR < 1$, then the graph is imbalanced. If $IR = 1$, then it is balanced.

\textbf{Definition 2.2. Cost Sensitive Learning.}
Cost-sensitive learning is kind of method which trains classifiers by giving different penalties for the misclassification of different classes of samples. Unlike traditional classification methods that reduce the misclassification rate as much as possible, cost-sensitive learning focuses on introducing different types of misclassification costs into the classification decision to reduce the overall cost of misclassification.

Without loss of generality, given a set of labels  $C$, there are a total of $K$ classes. The misclassification cost matrix $\mathbf{C}$ for multi-class classification problem can be defined in the following form:
 \begin{equation}
\mathbf{C}=\left[\begin{array}{cccc}
C_{10} & C_{11} & \cdots & C_{1 K} \\
C_{20} & C_{21} & \cdots & C_{2 K} \\
\vdots & \vdots & \ddots & \vdots \\
C_{K 0} & C_{K 1} & \cdots & C_{K K}
\end{array}\right]
\end{equation}
where $C_{ij}$ denotes the misclassification cost of classifying class  $i $ samples as class  $j $. The larger the value of $C_{ij}$, the larger the loss caused by this misclassification. $C_{ij} \subset[0,+\infty)$ is an associated cost item.

\textbf{Definition 2.3. GNN-based social network fraud detection.}
Taking the subscribers in the mobile network as nodes, subscribers' network behavior as node features, and the social connections between subscribers as edges, we can establish the social behavior graph 
$\mathcal{G}=(\mathcal{V},\mathcal{X},\mathbf{A},\mathcal{E},\mathcal{Y})$, where $\mathcal{V}=\{ v_1, v_2, v_3,...v_N \}$  is the set of nodes. $\mathcal{X}=\{\mathbf{x}_1,\mathbf{x}_2,...\mathbf{x}_N \}$  is the set of subscribers' original behavioral features, where $\mathbf{x}_i  \in  \mathbb{R} ^d$ is a behavioral features vector of subscriber $v_{i}$. Stacking these vectors into a matrix forms the feature matrix $\mathbf{x}_i  \in  \mathbb{R} ^d$ of the graph $\mathcal{G}$. $\mathbf{A} \in \mathbb{R}^{N \times N}$ denotes the adjacency matrix of the graph $\mathcal{G}$. $a_{i,j}  = 1$ indicates that there is an edge between node $i$ and node $ j$. If not, $a_{i,j}  = 0$. $\mathcal{Y} = \left \{y_{1}, y_{2},. . ,y_{n} \right \}$ is the set of labels. For a given mobile social network, graph neural networks can be used for node embedding learning. The inter-layer transfer formula in a GNN-based telecom fraud detection model can be formally described as:
\begin{equation}
\mathbf{h}_{v}^{(l)} =\sigma \left (  \mathbf{h}_{v}^{(l-1)}\operatorname{Agg} \left \{ \mathbf{h}_{v^{\prime}}^{(l-1)}:(v,v^{\prime})\in \mathcal{E}   \right \}   \right )
\end{equation}
where $\mathbf{h}_v^{(l)}$ represents the embedding of node $v$ at layer $l$, $\mathbf{h}_v^{(0)}=\mathbf{x}_v$, and $v^{\prime}$ is the neighbor of node $v$. $\operatorname{Agg}()$ is a message aggregation function for neighbor feature aggregation, such as mean aggregation, pooling aggregation, attention aggregation, etc. $\oplus$ represents feature concatenation or summation operation.

\section{Proposed model}\label{Method}

\subsection{Model Overview}
The complete pipeline of our proposed method is shown in Figure \ref{CSGNN}. First, we input the graph into a reinforcement learning-based neighbor sampler, and perform neighbor sampling on each node. The main idea is to train a suitable sampling threshold $p$ using reinforcement learning, and then perform top-$p$ sampling after sorting all of each node's neighbors according to their similarity to the central node. After obtaining the sampled graph, we input it into the cost-sensitive learner. To be specific, message aggregation is first conducted using GNN to obtain node embeddings. Subsequently, cost-sensitive training is performed using the improved cost-sensitive module so that the learner can obtain the optimal cost-sensitive matrix. Finally, the classification is achieved by using cost-sensitive node embedding representation.

\subsection{RL-based Neighbor Sampler}
Graph imbalance problem can significantly reduce GNN model performance. One of the reason is that positive nodes have too many negative neighbors (challenge 1 illustrated in Section \ref{introduction}). And the neighbor information aggregation by GNN can significantly interfere with the judgment of the model on the central node. In addition, the neighbor information has different contributions to the judgment of the central node, and even has interference sometimes \cite{2021GraphMix}. This challenge is fatal because GNN model always tends to discriminate the minority class as the majority class because this makes the loss smaller.  This inspires us to design a sampling mechanism to ensure that the adverse effect can be balanced. Specifically, we consider using reinforcement learning to sample the neighbors of each node, such that each central node retains only a fraction of its valid neighbors, thus facilitating unbiased GNN information aggregation. In implementation, we divide the reinforcement learning-based neighbor sampler into a node feature transformation module and a neighbor sampling module.

\subsubsection{Node feature transformation}
In general, we can sample the neighbors based on the similarity of the central and neighboring nodes. However, the appropriate way to measure the inter-node similarity has a large impact on the final effect. In prior works, AGCN \cite{li2018adaptive} employed Gaussian kernel normalized to the martingale distance. DIAL-GNN \cite{chen2019deep} applied parametric cosine similarity. And NSN \cite{liu2019neural} performed similarity measurement based on the inner product of bilinear similarity and hyperspherical learning strategy. The high time complexity of these methods, however, limits their ability to be applied to practical issues. These techniques can consume a large amount of time and computational resources due to the larger number of nodes and edges that real-world graph representation learning issues typically have. Inspired by GraphMix \cite{2021GraphMix} and CARE-GNN \cite{dou2020enhancing}, we design a parametric node similarity measure method. Specifically, we employ a fully connected network(FCN) as node feature translator in the proposed model. And we use the Euclidean distance between the predictions of two nodes as a node similarity measure. For the feature $\mathbf{x}_{v}$ of node $v$ , the transformed embedding $\mathbf{h}_{v}$ is represented as follows:
\begin{equation}
\mathbf{h}_{v}=\sigma \left ( \mathbf{W} \mathbf{x}_{v} \right ) 
\label{trans_feature}
\end{equation}
where $\sigma$ is the activation function, ReLU is used in impletation, and $\mathbf{W} $ is the network weights to be learned.

When calculating the distance between an intermediate node $v$ and its neighboring nodes $v^{\prime}$, we can take its transformed embedding $h^{(l)}$ as input and use the Euclidean distance to calculate the following.
\begin{equation}
\mathcal{D}\left(v, v^{\prime}\right)=\left\|\mathbf{h}_{v}-\mathbf{h}_{v^{\prime}}\right\|_{2}
\end{equation}

According to the distance $\mathcal{D}$, we can calculate the similarity between the neighboring nodes $v, v^{\prime}$ as:
\begin{equation}
S\left(v, v^{\prime}\right)=1-\mathcal{D}\left(v, v^{\prime}\right)
\label{similarity}
\end{equation}

Our objective is to make the transformed node features closer together in high-dimensional space and farther apart for different class. To train the fully connected layer using signals from the labels, we use the minimized cross-entropy loss function as the objective of the optimization.
\begin{equation}
\mathcal{L}_{trans} = -\sum_{v \in V} \mathbf{y}_{v} \log \left(\mathbf{h}_{v}\right)
\label{trans_loss}
\end{equation}

\subsubsection{Node neighbor sampling}
When perform sampling, we use reinforcement learning algorithm to design an adaptive sampling criterion to automatically sample the optimal number of similar neighbors. Specifically, based on the neighbor similarity learning in the previous step, we use top-$p$ sampling to filter out similar neighbors that match the requirements. To determine the optimal threshold $p$, we design a reinforcement learning (RL) algorithm based on Bernoulli's multi-armed bandit (BMAB). After finding the optimal threshold $p$, we rank the similarity of neighboring nodes and the central node. Then, for each node $v$, we sample its $p*\left | N(v) \right | $ neighbors. The whole process is described in detail below.

The BMAB model can be expressed as B($\mathcal{S}$, $\mathcal{A}$, f , T ), where $\mathcal{S}$ is the state space, $\mathcal{A}$ is the action space, f is the reward function, and T is the termination condition. Given an initial sampling probability$p$, the Agent chooses an action based on the current state. The reward depends on the difference in the average similarity between two consecutive epochs.

\noindent\textbf{State:}
In order to illustrate the state space of BMAB, we first define the average similarity of the training set nodes in CSGNN of the $e$-th epoch as:
\begin{equation}
G\left(S\right)^{(e)}=\frac{\sum_{v, \in \mathcal{V}_{\text {train }}} S\left(v, v^{\prime}\right)^{(e)}}{\left|\mathcal{E}^{(e)}\right|}
\label{average similarity}
\end{equation}
where $\mathcal{E}^{(e)}$ is set of edges in the training set $\mathcal{V}$.

We define the positive and negative difference of the average similarity of the nodes of two adjacent epochs as the BMAB state. That is, the state set  $ \mathcal{S}=\{ s_{1}, s_{2}\} $, where state $s_{1}$ denotes that the average node similarity under the current epoch is greater than that of the previous epoch, i.e., $G\left(S\right)^{(e-1)}<G\left(\mathcal{S}\right)^{(e)}$, and  $s_{2}$  denotes that the average node similarity under the current epoch is less than or equal to that of the previous epoch, i.e., $G\left(S\right)^{(e-1)} \geqslant G\left(S\right)^{(e)}$.

\noindent\textbf{Action:}
The objective of the reinforcement learning module is to learn the optimal neighbor filtering threshold $p$. During the running of the model, BMAB is required to adjust $p$ according to the real-time state $s$. Here, we give the action space  $\mathcal{A} = \{ a_{1}, a_{2}\}$, where action  $a_{1}$: gives the sampling probability $p$ plus a fixed step  $ \tau$, and $a_{2}$: gives the sampling probability $p$ minus a fixed step $ \tau$.  For any state $s \in  \mathcal{S}$,  we define the mapping relation $ \pi$  between states and actions as follows.
\begin{equation}
a=\pi(s)=\left\{\begin{array}{ll}
p+\tau, & s_{2} \\
p-\tau, & s_{1}
\end{array}\right.
\label{action}
\end{equation}
where $ \tau$ is the change step of the neighbor sampling probability $p$. It's set manually as a hyperparameter of the model.

\noindent\textbf{Reward:}
A suitable sampling probability $p$ should guarantee that the optimal number of similar neighbor nodes are sampled. Therefore, we design the reward mechanism to give a positive reward if the Agent makes the average similarity of nodes in the training set increase under $p$, and a negative reward otherwise, which is formally described as follows:
\begin{equation}
f(p , a)^{(e)}=\left\{\begin{array}{l}
+1, \quad G\left(S\right)^{(e-1)} \leq G\left(S\right)^{(e)} \\
-1, \quad G\left(S\right)^{(e-1)}>G\left(S\right)^{(e)}
\end{array}\right.
\label{award&penalty}
\end{equation}

The reward is positive when the average distance of epoch $e$ newly selected neighbor is smaller than the previous epoch and vice versa. Estimating the cumulative reward is not easy. Therefore, we use immediate rewards to update actions without exploration greedily. Specifically, we increase $p$ with a positive reward and vice versa.

\noindent\textbf{Termination:}
A proper termination condition is crucial for BMAB to achieve the optimal threshold $p$ and save computational resources. If it converges in the last 15 epochs, we think that an optimal threshold $p^{(l)}$  is found, and the RL module is terminated. After termination, the filtering threshold is fixed to the optimal value until the GNN converges:
\begin{equation}
\left|\sum_{e-15}^{e} f\left(p, a\right)^{(e)}\right| \leq 2, \text { where } e \geq 15
\label{Termination}
\end{equation}

\subsection{Cost-Sensitive Learner}
After performing neighbor sampling, the new graph can significantly reduce the number of negative neighbor nodes of positive nodes. However, we should not reduce this number indefinitely. An extreme example is to balance the number of positive and negative samples in the graph completely. But this situation is also damaging to the model, because an excessive number of negative nodes being filtered out leads to a large amount of valid data loss (challenge 2 illustrated in Section \ref{introduction}). To solve this problem, we use a well-designed cost-sensitive learner for training GNN output embeddings after performing the neighbor sampling. The designed cost-sensitive learner contains two parts. One is the GNN node embedding learning module, and the other is the cost-sensitive module.

\subsubsection{Neighbor information aggregation}
Previous works tend to use attention mechanisms \cite{liu2019geniepath, wang2019semi} or design weighting parameters \cite{liu2018heterogeneous} for message aggregation. This is the current work done by many popular GNN models who improve on the message passing and aggregation mechanisms to enhance GNN performance. However, assuming that we have selected the most similar neighbors for each node, the attention coefficient or weighting parameter has no obvious meaning in neighbor information aggregation. On the contrary, it will increase the computational cost significantly. Therefore, in order to save computational costs while retaining important information about the domain, we use a simple and efficient aggregator, the average aggregator in GraphSAGE \cite{hamilton2017inductive}, for message aggregation. Compared with the attention mechanism, this method significantly reduces the model complexity. The formal description is as follows:
\begin{equation}
\mathbf{h}_{v}^{(l)}=\sigma\left(\mathbf{W} \cdot AGG\left(\left\{\mathbf{h}_{v}^{(l-1)}\right\} \cup\left\{\mathbf{h}_{v^{\prime}}^{(l-1)}, \forall v^{\prime} \in \mathcal{N}(v)\right\}\right)\right.
\label{GNN_embedding}
\end{equation}
where $\sigma$ is the activation function, $\mathbf{W}$ is the weight to be learned, and $\mathcal{N}(v)$ is the neighborhood of node $v$. In implementation , we use $MEAN()$ as aggregation function $AGG()$. $MEAN()$ denotes the operation of averaging the embeddings of the nodes in the set.  By noting the node embedding $\mathbf{h}_{v}^{(l)}$ as $\mathbf{z}_{v}$ for the last layer of GNN output, we can perform downstream tasks.

\subsubsection{Cost-sensitive learning} 
\label {Cost-sensitive learning module}
Ordinarily, it is feasible to run downstream tasks, such as node classification, after obtaining the node embedding $\mathbf{z}_{v}$. However, in an unbalanced graph, such an operation usually does not give the desired result. To tackle this problem, we try to introduce cost-sensitive learning and perform the cost-sensitive transformation on the output embeddings of GNN. Inspired by Salman H et al.\cite{khan2017cost}, we transform $\mathbf{z}_{v}$ using a cost-sensitive matrix to obtain the probability that node $v$ belongs to $k$.
\begin{align}
p_{k}(v) & = \operatorname{Softmax}\left(\mathbf{C}[v, k] \cdot  \mathbf{z}_{v}[k]\right)  \notag
 \\ 
& = \frac{\mathbf{C}_{v, k} \exp \left(\mathbf{z}_{v}[k]\right)}{\sum_{k^{\prime}} \mathbf{C}_{v, k^{\prime}} \exp \left(\mathbf{z}_{v}\left[k^{\prime}\right]\right)}
\label{pkv}
\end{align}
where $\mathbf{C}[v, k]$ is the value in the cost matrix $\mathbf{C}$ as defined in Section \ref{Preliminary}, and $\mathbf{z}_{v}[k]$ is the $k$-th dimensional value of $\mathbf{z}_{v}$ in the final embedding of the GNN.

However, a major challenge in achieving the above goal is how to reasonably determine the value of each element in the cost matrix $\mathbf{C}$. In the past work, cost-sensitive matrix has been given according to expert experience, or imbalance rate in the given data. Unfortunately, expert experience is costly and even difficult to obtain for many problems. If calculated according to the imbalance rate, these calculations cannot automatically adapt to different data sets when the data distribution changes, which makes it difficult to obtain optimal results. To resolve this problem, we design a cost-sensitive matrix algorithm that can be automatically optimized.

We consider that the optimal cost matrix should take into account the distribution matrix of different classes of data in the training set, the scatter matrix of embeddings, and the confusion matrix of classification results. For the data distribution matrix, the majority class samples should be assigned a lower cost, while the minority class sample should be the reverse. For the embedding scatter matrix, embeddings of samples of the same class should have smaller distances in the high-dimensional space, and samples of different classes should have larger distances. For the confusion matrix, misclassified samples in the current epoch should be given more weight in the next epoch, and the correct samples should be reversed.

Based on the above idea, we define the data distribution histogram matrix $\mathbf{H}$ as:
\begin{equation}
\mathbf{H}(i, j) = \left\{\begin{array}{ll}
\max \left(h_{i}, h_{j}\right): & i \neq j,(i, j) \in C \\
h_{i}: & i = j, i \in C
\end{array}\right.
\label{H_matrix}
\end{equation}
where $h_{i}$ denotes the histogram distribution of class $i$. $C$ is the node class in set $\left \{ C_{1}, . . ,C_{K} \right \}$.

To measure the distance of node embedding in high dimensional space, we use the scatter matrix to calculate within-class and between-class distances. We define Within-class scatter matrix ($\mathbf{S}_{W}$) and Between-class scatter matrix ($\mathbf{S}_{B}$), and Total scatter matrix ($\mathbf{S}$).  $\mathbf{S}_{W}$ is defined as follows:
\begin{equation}
\mathbf{S}_{W}=\sum_{i=1}^{K}\left(\frac{1}{N_{i}} \sum_{j=1}^{N_{i}}\left(\mathbf{x}_{i, j}-\mathbf{m}_{i}\right)\left(\mathbf{x}_{i, j}-\mathbf{m}_{i}\right)^{\top}\right)
\end{equation}
where $\mathbf{x}_{i, j}$ denotes the embedding of the $j$-th sample of class $i$ and $m_{i}$ denotes the average embedding of the samples of class $i$.
The within-class scatter matrix $\mathbf{S}_{B}$ is defined as follows:
\begin{equation}
\mathbf{S}_{B}=\sum_{i=1}^{K}\left(\mathbf{m}_{i}-\mathbf{m}\right)\left(\mathbf{m}_{i}-\mathbf{m}\right)^{\top}
\end{equation}

We always expect that $\mathbf{S}_{W}$ learned by the model is small and $\mathbf{S}_{B}$ is large. Thus, the smaller each value in the $\mathbf{S}$ matrix, the more effective the classifier is. In this way, we can define the overall scatter matrix as follows:
\begin{equation}
\mathbf{S}=\frac{\mathbf{S}_{W}}{\mathbf{S}_{B}}
\label{S_matrix}
\end{equation}

We use the normalized confusion matrix as $\mathbf{R}$. The elements in $\mathbf{R}$ are defined as follows.
\begin{equation}
\mathbf{R}(i, j)=\frac{\left|\left\{v \mid \underset{k}{\operatorname{argmax}}  \\\  p_{k}(v)=j, y_{v}=i\right\}\right|}{|\mathcal{V}|}
\label{R_matrix}
\end{equation}
where $p_{k}(v)$ is defined in Eq.\ref{pkv}.

After obtaining the three matrices $\mathbf{H}$,  $\mathbf{S}$, and $\mathbf{R}$, we consider the synthesis of them three as the training objective $\mathbf{T}$ of the cost matrix $\mathbf{C}$, which is defined as follows:
\begin{equation}
\mathbf{T}=\beta \cdot \mathbf{H} \odot \mathbf{S} \odot \mathbf{R}
\label{T_matrix}
\end{equation}
where $\beta$ is the a hyperparameter, and $\odot$ is matrix Hadamard product.

With the cost matrix training target $\mathbf{T}$, we can control the training process. To properly measure the disparity between the trained cost matrix $\mathbf{C}$ and the target $\mathbf{T}$, we define the loss between the two as $\mathcal{L}_{\text {cost }}(\mathbf{C}, \mathbf{T})$.
\begin{equation}
\mathcal{L}_{\text {cost }}(\mathbf{C},\mathbf{T})=\|\mathbf{T}-\mathbf{C}\|_{2}^{2}+E_{v a l}(\theta, \mathbf{C})
\label{loss_cost}
\end{equation}
where $E_{v a l}()$ is the validation error under current parameters.

Another challenge that still exists is how to initialize the value of the cost matrix $\mathbf{C}$. A natural idea is to adopt popular initialization methods for neural network weights, such as Xavier initialization, Kaiming initialization, and random initialization. However, these initialization methods are not always applicable to cost-sensitive matrices. On the one hand, the size of $\mathbf{C}$ is usually small ($K*K$ dimensions) , which is much smaller than the number of neurons in a neural network. It is not enough on a small number of elements to demonstrate the effectiveness of randomization methods that are only effective on a large number of elements. On the other hand, these initialization methods are unable to capture the original objective of cost-sensitive learning, i.e., the corrective effect of cost on the minority samples in the model. Therefore, the above initialization techniques are of no apparent significance for the learning of the cost matrix. To solve this problem, we start from the initial goal of the cost matrix, which is to solve the imbalance problem in the sample. Meanwhile, we learn from the idea of pre-training in machine learning and consider assigning $\mathbf{C}$ an excellent initial value that contains the imbalance rate, then execute fine-tuning in the future training phase. Specifically, we start from the imbalance rate $IR$ of the sample and make a $\log 1p$ transformation of the inverse of IR, which can make the biased cost values in the initial cost matrix conform to an unbiased normal distribution.
\begin{equation}
\mathbf{C} _{i j}=\log 1p\left(\frac{1}{I R}\right)=\log \left(\frac{\left|C _{j}\right|}{\left|C _{i}\right|}+1\right)
\label{C_initialization}
\end{equation}

After obtaining the initialized value of the cost matrix, we optimize $\mathcal{L}_{\text {cost }}(\mathbf{C},\mathbf{T})$ at each subsequent epoch according to a certain learning rate. The optimal cost matrix $\mathbf{C}^{*}$ can be expressed as follows:
\begin{equation}
\mathbf{C}^{*}=\underset{\mathbf{C}}{\operatorname{argmin}} \\\ \mathcal{L}_{\text {cost }}(\mathbf{C},\mathbf{T})
\end{equation}

To optimize the cost function in the above equation, we use the gradient descent (GD) algorithm, which computes the direction of the update step as follows:
\begin{equation}
\begin{aligned}
\nabla \mathcal{L}_{\text {cost }}(\mathbf{C},\mathbf{T}) &=\nabla\left(\mathbf{v}_{a}-\mathbf{v}_{b}\right)\left(\mathbf{v}_{a}-\mathbf{v}_{b}\right)^{T} \\
&=\left(\mathbf{v}_{a}-\mathbf{v}_{b}\right){\mathbf{J}_{\mathbf{v}_{b}}^{T}} \\
&=-\left(\mathbf{v}_{a}-\mathbf{v}_{b}\right) \mathbf{1}^{T}
\end{aligned}
\end{equation}
where $\mathbf{v}_{a}=\operatorname{vec}(\mathbf{T}), \mathbf{v}_{b}=\operatorname{vec}(\mathbf{C})$, $\mathbf{J}$ denotes the Jacobi matrix and $\mathbf{1}$ denotes the unit matrix.

The above design can automatically optimize the cost matrix $\mathbf{C}$. However, our approach modifies the output embedding in the GNN training process (Eq.\ref{pkv}). Therefore, this has the potential to impact on the subsequent loss computation and backpropagation. We will discuss their impact on the backpropagation algorithm and theoretically demonstrate the feasibility of such cost-sensitive loss classification in the next subsection.

\subsection{Proof of Cost-Sensitive Design}\label{proof}
To demonstrate that the cost-sensitive design can bring classification improvement to GNN, we refer to \cite{khan2017cost,bartlett2006convexity,beijbom2014guess} to give two proposition proofs. \textbf{Proposition 1} demonstrates that the loss function of the cost-sensitive design is available for classification, and \textbf{Proposition 2} demonstrates that the cost-sensitive design does not affect the computation of the gradient in the back-propagation process of the GNN model.

\begin{Proposition}
The modified cost-sensitive loss function $\mathcal{L}_{GNN}$ is classification calibrated (c-calibrated).
\begin{equation}
\mathcal{L}_{GNN} = -\sum_{v \in V} \mathbf{y}_{v} \log \left(\mathbf{p}_{v}\right)
\label{GNN_loss}
\end{equation}
where 
\begin{equation}
\mathbf{p}_v[k]=p_{k}(v)= \frac{\mathbf{C}_{v, k} \exp \left(\mathbf{z}_{v}[k]\right)}{\sum_{k^{\prime}} \mathbf{C}_{v, k^{\prime}} \exp \left(\mathbf{z}_{v}\left[k^{\prime}\right]\right)}
\end{equation}

\begin{proof}
For a given node $v$, its one-hot label vector is $\mathbf{y}_{v}$. The modified cost-sensitive cross-entropy loss function is:
\begin{equation}
\mathcal{L}\left(\mathbf{C},\mathbf{y}_{v}, \mathbf{z}_{v}\right)=-\mathbf{y}_{v}\log \left(\frac{\mathbf{C}_{v, k} \exp \left(\mathbf{z}_{v}[k]\right)}{\sum_{k^{\prime}} \mathbf{C}_{v, k^{\prime}} \exp \left(\mathbf{z}_{v}\left[k^{\prime}\right]\right)}\right)
\end{equation}
According to the sample data distribution and the conditional probability $p\left(y_{v} \mid \mathbf{x}_{v}\right)$,  we can get the classification risk:
\begin{equation}
\begin{aligned}
R_{l}\left(\mathbf{z}_{v}\right) &=\mathbb{E}_{\mathbf{x}_{v} \sim D}\left[\mathcal{L}\left(\mathbf{C},\mathbf{y}_{v}, \mathbf{z}_{v}\right)\right] \\
&=\sum_{y=1}^{K} p\left(y \mid \mathbf{x}_{v}\right) \mathcal{L}\left(\mathbf{C},\mathbf{y}_{v}, \mathbf{z}_{v}\right) \\
&=-\sum_{y=1}^{K} p\left(y \mid \mathbf{x}_{v}\right) \log \left(\frac{\mathbf{C}_{v, k} \exp \left(\mathbf{z}_{v}[k]\right)}{\sum_{k^{\prime}} \mathbf{C}_{v, k^{\prime}} \exp \left(\mathbf{z}_{v}\left[k^{\prime}\right]\right)}\right)
\end{aligned}
\end{equation}
To find the optimal solution, we take the derivative of the above equation.
\begin{equation}
\frac{\partial R_{l}\left(\mathbf{z}_{v}\right)}{\partial z_{y v}}=\left.\frac{\partial R_{l}\left(z_{y}\right)}{\partial z_{y_{v}}}\right|_{y \neq y_{v}}+\left.\frac{\partial R_{l}\left(z_{y}\right)}{\partial z_{y_v}}\right|_{y=y_{v}}
\end{equation}
When $y=y_{v}$,
\begin{equation}
\begin{aligned}
&\left.\frac{\partial R_{L}\left(z_{y}\right)}{\partial z_{y_{v}}}\right|_{y=y_{v}}=\\
& \frac{\partial}{\partial z_{y_{v}}}\left(-\sum_{y=1}^{K} p\left(y \mid \mathbf{x}_{v}\right) \log \left(\mathbf{C}_{y_{v}, y_{v}} \exp \left(\mathbf{z}_{v}\left[y_{v}\right]\right)\right)\right.\\
&+\sum_{y_{v}=1}^{K} p\left(y \mid \mathbf{x}_{v}\right) \log \left(\sum_{k^{\prime}} \mathbf{C}_{y_{v}, k^{\prime}} \exp \left(\mathbf{z}_{v}\left[k^{\prime}\right]\right)\right) \\
&=-p\left(y_{v} \mid \mathbf{x}_{v}\right) \frac{\mathbf{C}_{y_{v}, y_{v}} \exp \left(\mathbf{z}_{v}\left[y_{v}\right]\right)}{\mathbf{C}_{y_{v}, y_{v}} \exp \left(\mathbf{z}_{v}\left[y_{v}\right]\right)} \\
&+p\left(y_{v} \mid \mathbf{x}_{v}\right) \frac{\mathbf{C}_{y_{v}, y_{v}}  \exp \left(\mathbf{z}_{v}\left[y_{v}\right]\right)}{\sum_{k^{\prime}} \mathbf{C}_{y_v}, k^{\prime} \exp \left(\mathbf{z}_{v}\left[k^{\prime}\right]\right)} \\
&= p\left(y_{v} \mid \mathbf{x}_{v}\right) \frac{C_{y_{v}, y_{v}}  \exp \left(\mathbf{z}_{v}\left[y_{v}\right]\right)}{\sum_{k^{\prime}} C_{y_{v}, k^{\prime}}  \exp \left(\mathbf{z}_{v}\left[k^{\prime}\right]\right)}-p\left(y_{v} \mid \mathbf{x}_{v}\right)
\end{aligned}
\end{equation}
When $y \neq y_{v}$:
\begin{equation}
\begin{aligned}
&\frac{\partial R_{l}\left(z_{y}\right)}{\partial z_{y_{v}}}C_{y \neq y_{v}}=\\
& \frac{\partial}{\partial z_{y_{v}}}\left(-\sum_{y=1}^{k} p\left(y \mid \mathbf{x}_{v}\right) \log \left(\mathbf{C}_{y, y} \exp \left(\mathbf{z}_{v}[y]\right)\right)\right)\\
&+\sum_{y=1}^{k} p\left(y \mid \mathbf{x}_{v}\right) \log \left(\sum_{k^{\prime}} \mathbf{C}_{y, k^{\prime}} \exp \left(\mathbf{z}_{v}\left[k^{\prime}\right]\right)\right) \\
&= \sum_{y \neq y_{v}} p\left(y \mid \mathbf{x}_{v}\right) \cdot \frac{C_{y, y_{v}} \exp \left(\mathbf{z}_{v}\left[y_{v}\right]\right)}{\sum_{k^{\prime}} C_{y, k^{\prime}} \exp \left(\mathbf{z}_{v}\left[k^{\prime}\right]\right)}
\end{aligned}
\end{equation}
By observing the above two situations, we can combine the two expressions and get:
\begin{equation}
\frac{\partial R_{l}\left(z_{y}\right)}{\partial z_{y_{v}}}=p\left(y \mid \mathbf{x}_{v}\right) \frac{\mathbf{C}_{y, y_{v}} \exp \left(\mathbf{z}_{v}\left[y_{v}\right]\right)}{\sum_{k^{\prime}} C_{y, k^{\prime}} \exp \left(\mathbf{z}_{v}\left[k^{\prime}\right]\right)}-p\left(y_{v} \mid \mathbf{x}_{v}\right)
\label{derivative}
\end{equation}
To find the optimal solution, let $\frac{\partial R_{l}\left(z_{y}\right)}{\partial z_{y_{v}}}=0$, we can get:
\begin{equation}
\exp \left(\mathbf{z}_{v}\left[y_{v}\right]\right)=\frac{p\left(y_{v} \mid \mathbf{x}_{v}\right) \sum_{k^{\prime}} \mathbf{C}_{y_{\cdot} k^{\prime}} \exp \left(\mathbf{z}_{v}\left[k^{\prime}\right]\right)}{p\left(y \mid \mathbf{x}_{v}\right) \mathbf{C}_{y, y_{v}}}
\end{equation}
By taking the logarithm of both sides of the above equation at the same time yields
\begin{equation}
\begin{aligned}
\mathbf{z}_{v}\left[y_{v}\right]=& \log p\left(y_{v} \mid \mathbf{x}_{v}\right)+\log \left(\sum_{k\prime} \mathbf{C}_{y,k^{\prime}} \exp \left(\mathbf{z}_{v}\left[k^{\prime}\right]\right)\right)\\
&-\log \left(p\left(y \mid \mathbf{x}_{v}\right)\mathbf{C}_{y, y_{v}}\right)
\end{aligned}
\label{zvyv}
\end{equation}
The model output for cost-sensitive design and the Bayesian cost of class $y$ are shown to have an inverse connection, as seen by the negative term $-\log \left(p\left(y \mid \mathbf{x}_{v}\right)\mathbf{C}_{y, y_{v}}\right)$ in Eq.(\ref{zvyv}), so the cost loss is classification calibration. Additionally, the extreme value is only achieved when $y=y_{v}$ which shows that the design is guess aversion, thus the modified loss function is suitable for classification.
\end{proof}

\end{Proposition}

\begin{Proposition}
The gradient calculation during GNN back propagation is unaffected by the modified cost-sensitive loss function.

\begin{proof}
We begin by calculating the partial derivative of a softmax neuron with respect to its input.
\begin{equation}
\frac{\partial p_k(v)}{\partial \mathbf{z}_{v}[m]}=\frac{\partial}{\partial \mathbf{z}_{v}[m]}\left(\frac{\mathbf{C}_{v,k} \exp \left(\mathbf{z}_{v}[k]\right)}{\sum_{k\prime} \mathbf{C}_{v, k^{\prime}} \exp \left(\mathbf{z}_{v}\left[k^{\prime}\right]\right)}\right)
\end{equation}
There are two cases, $m = k$ or $m \ne k$. We first solve for the case $m = k$:
\begin{align}
\frac{\partial p_k(v)}{\partial \mathbf{z}_{v}[m]}=& \frac{\mathbf{C}_{v,m}}{\left(\sum_{k\prime} \mathbf{C}_{v, k^{\prime}} \exp \left(\mathbf{z}_{v}\left[k^{\prime}\right]\right)\right)^{2}} \nonumber\\
&\cdot \bigg(\exp \left(\mathbf{z}_{v}[m]\right) \sum_{k} \mathbf{C}_{v, k^{\prime}} \exp \left(\mathbf{z}_{v}\left[k^{\prime}\right]\right) \bigg. \nonumber\\
&\bigg. -\mathbf{C}_{v,m} \exp \left(2 \mathbf{z}_{v}[m]\right)\bigg)
\end{align}
After simplification, we get :
\begin{equation}
\frac{\partial p_k(v)}{\partial \mathbf{z}_{v}[m]}=p_m(v)\left(1-p_m(v)\right), \quad \text { s.t.: } m=k
\end{equation}
Next, we solve for the case when $m \neq k$.
\begin{align}
\frac{\partial p_k(v)}{\partial \mathbf{z}_{v}[m]}&=-\frac{\mathbf{C}_{v,k} \mathbf{C}_{v,k} \exp \left(\mathbf{z}_{v}[m]+\mathbf{z}_{v}[k]\right)}{\left(\sum_{k\prime} \mathbf{C}_{v, k^{\prime}} \exp \left(\mathbf{z}_{v}\left[k^{\prime}\right]\right)\right)^{2}}  \nonumber \\
&=-p_m(v) p_k(v), \quad \text { s. t.: } m \neq k
\end{align}
The loss function can be differentiated as follows:
\begin{align}
\frac{\partial \ell(\mathbf{y}, \mathbf{d})}{\partial \mathbf{z}_{v}[m]} &=-\sum_{n} d_{n} \frac{1}{p_k(v)} \frac{\partial p_k(v)}{\partial \mathbf{z}_{v}[m]} \nonumber\\
&=-d_{m}\left(1-p_m(v)\right)+\sum_{n \neq m} d_{n} p_m(v) \nonumber\\
&=-d_{m}+\sum_{n} d_{n} p_m(v)
\end{align}
Because $d$ is defined as the distribution of the model output over all classes, we have $\sum_{n} d_{n}=1$. Therefore:
\begin{equation}
\frac{\partial \mathcal{L}\left(\mathbf{C},\mathbf{y}_{v}, \mathbf{z}_{v}\right)}{\partial \mathbf{z}_{v}[m]}=-d_{m}+p_m(v)
\end{equation}
This result is the same as the case where the vanilla cross-entropy loss function does not contain any cost-sensitive parameters. Therefore, we can conclude that although the cost-sensitive design in this paper changes the GNN output from $\mathbf{z}_v$ to $p_k(v)$, the gradient formulation remains unchanged and does not affect the back propagation process.
\end{proof}

\end{Proposition}

\subsection{The proposed algorithm}
Given an imbalanced fraud detection graph $\mathcal{G}$ and a node in $\mathcal{V}_{train}$, its cost-sensitive embedding is $\mathbf{p}(v)$ . We can define the CE loss of GNN module as Eq.(\ref{GNN_loss}):
\begin{equation}
\mathcal{L}_{GNN} = -\sum_{v \in V} \mathbf{y}_{v} \log \left(\mathbf{p}_{v}\right)
\label{GNN_loss_final}
\end{equation}

Together with the loss function of the node feature transformation in Eq.(\ref{trans_loss}), we define the loss of CSGNN as:
\begin{equation}
\mathcal{L}_{CSGNN}=\mathcal{L}_{GNN}+\lambda \mathcal{L}_{trans}
\label{model loss}
\end{equation}
where $\lambda$ is the weighting parameter.

The complete train processing of the proposed CSGNN is shown in Algorithm \ref{algorithm}.

 \IncMargin{1em}
\begin{algorithm} [ht] \SetKwData{Left}{left}\SetKwData{This}{this}\SetKwData{Up}{up} \SetKwFunction{Union}{Union}\SetKwFunction{FindCompress}{FindCompress} \SetKwInOut{Input}{input}\SetKwInOut{Output}{output}

	\KwIn{An undirected graph with node features and labels: $\mathcal{G}=(\mathcal{V},\mathcal{X},\mathbf{A},\mathcal{E},\mathcal{Y}))$ \;
	\quad \quad \quad The number of node classes $K$\;
	\quad \quad \quad Number of layers, epochs: $L,E$\\
	  }

	\KwOut{Vector representations $\mathbf{p}_v$, $v \in \mathcal{V}_{train}$.}

	 \BlankLine 
	 \ \tcp*[h]{Initialization}\\
	 \ $ \mathbf{C}_{ij}\leftarrow$ Eq.(\ref{C_initialization}), $ {\forall} v \in \mathcal{V}_{train}$ \\
	 \ \tcp*[h]{ Train the module}\\
     \For {$e = 1,2...E$}
     {  
        \ \tcp*[h]{Feature transformation }\\
        \ $\mathbf{h}_v \leftarrow$ Eq.(\ref{trans_feature}) \;
        \ $\mathcal{L}_{trans} \leftarrow $ Eq.(\ref{trans_loss})
        
        \ \tcp*[h]{RL-based neighbor sampler }\\
        \While{Eq.(\ref{Termination}) is False}
        {
				\ $ S\left(v, v^{\prime}\right)   \leftarrow $ Eq.(\ref{similarity}) \;
				\ $ G\left(S\right)^{(e)} \leftarrow $ Eq.(\ref{average similarity})\;
				\ $a \leftarrow$ Eq.(\ref{action})\;
				\ $f(p,a)^{(e)} \leftarrow$ Eq.(\ref{award&penalty})
        }
      \ \tcp*[h]{Graph neural network }\\

		 	\For{$l=1,2,...L$}
		 	{
         \ $\mathbf{h}_{v}^{(l)} \leftarrow $ Eq.(\ref{GNN_embedding})  \;
		    }
		 \ $\mathbf{z}_{v} \leftarrow $ $\mathbf{h}_{v}^{(L)}$  \tcp*{GNN output embedding}
		 \ $p_{k}(v) \leftarrow $ Eq.(\ref{pkv})  \tcp*{Cost-sensitive embedding}
	   \ $\mathcal{L}_{GNN} \leftarrow$ Eq.(\ref{GNN_loss_final})  \tcp*{GNN loss}
	   \ $\mathbf{H} \leftarrow$ Eq.(\ref{H_matrix}), $\mathbf{S} \leftarrow$ Eq.(\ref{S_matrix}), $\mathbf{R} \leftarrow$ Eq.(\ref{R_matrix}) \;
	   \ $\mathbf{T} \leftarrow$ Eq.(\ref{T_matrix})  \tcp*{Cost matrix target}
		 \ $\mathcal{L}_{cost} \leftarrow$ Eq.(\ref{loss_cost})  \tcp*{Cost matrix loss}
		 \ $\mathcal{L}_{CSGNN} \leftarrow$ Eq.(\ref{model loss})  \tcp*{CSGNN loss}
     }
     
 	 	  \caption{Cost-Sensitive Graph Neural Network}
 	 	  \label{algorithm} 
 	 \end{algorithm}
 \DecMargin{1em}

\section{EXPERIMENTS}

This section will present the performance of CSGNN on the imbalanced telecom fraud detection dataset and comparisons with the baseline methods. It mainly answers the following research questions.
\begin{itemize}
\item RQ1: Does CSGNN outperform the state-of-the-art methods for graph-based anomaly detection?
\item RQ2: How does CSGNN perform with respect to the graph imbalance problem?
\item RQ3: How do the key components of CSGNN contribute to the final classification?
\item RQ4: What is the hyperparameter sensitivity and its impact on model design?
\end{itemize}

\subsection{Experimental setup}
\subsubsection{Datasets}
In this paper, we use two imbalanced telecom fraud detection datasets from the real world, namely Sichuan and BUPT.

Sichuan: The dataset contains the CDR data of 6,106 users from 23 cities in a large province of China, spanning from August 2019 to March 2020 \cite{hu2022btg}, where the node features are extracted from the features of user behavior records, and the edges are constructed by Hu et al. \cite{hu2022btg} based on the similarity of the node features. All samples contain two classes: fraudsters and non-fraudsters. In this dataset, the imbalance rate $IR=1962/4144=0.4735$.

BUPT: The dataset includes CDR data for one week from users in a Chinese city \cite{liu2019agrm}. The author extracted features from the raw CDR data and created connected edges based on user communication behavior. The dataset contains 3 categories, where the data imbalance rate $IR=8074/99861=0.0809$, and the details are shown in the following table.

\begin{table}[ht]
\centering
\caption{Dataset and graph statistics.}
\renewcommand{\arraystretch}{1.5}
\setlength{\tabcolsep}{1.6mm}{
\begin{tabular}{|c|c|c|c|c|c|}
\hline
\textbf{Dataset} &
  \begin{tabular}[c]{@{}c@{}}\textbf{Nodes}\\ \textbf{(fraud ratio)}\end{tabular} &
  \begin{tabular}[c]{@{}c@{}}\textbf{Edges}\end{tabular} &
  \begin{tabular}[c]{@{}c@{}}\textbf{Classes}\end{tabular} &
  \begin{tabular}[c]{@{}c@{}}\textbf{Features}\end{tabular} &
  \textbf{IR} \\ \hline
\multirow{2}{*}{\textbf{Sichuan}} & \multirow{2}{*}{6106(32.1\%)} & \multirow{2}{*}{838528} & Benign:4144     & \multirow{2}{*}{55} & \multirow{2}{*}{0.4735} \\ \cline{4-4}
                                  &                               &                         & Fraud:1962 &                     &                         \\ \hline
\multirow{3}{*}{\textbf{BUPT}}    & \multirow{3}{*}{116,383 (7.3\%)}               & \multirow{3}{*}{350751}  & Normal: 99861                 & \multirow{3}{*}{39}        & \multirow{3}{*}{0.0809} \\ \cline{4-4}
                                  &                                                &                          & Fraudster : 8448              &                            &                         \\ \cline{4-4}
                                  &                                                &                          & Courier : 8074                &                            &                         \\ \hline
\end{tabular}}
\end{table}

\begin{table*}[ht]
\caption{Performance comparison on two real-world social network fraud datasets.}
\renewcommand{\arraystretch}{1.7}
\setlength{\tabcolsep}{1.0mm}{
\begin{tabular}{|c|l|lll|llll|llll|ll|l|}
\hline
\multicolumn{1}{|l|}{\textbf{Datasets}} & \textbf{Method}                           & \multicolumn{3}{c|}{\textbf{General GNNs}}                                                                                            & \multicolumn{4}{c|}{\textbf{GNN-based Fraud Detector}}                                                                                                                                                                           & \multicolumn{4}{c|}{\textbf{Imbalance Graph Learner}}                                                                                                                                                                                                                                                                    & \multicolumn{2}{c|}{\textbf{Ablation}}                 & \multicolumn{1}{c|}{\textbf{Ours}} \\ \hline
                                        & \textbf{Metric}                           & \multicolumn{1}{l|}{\textbf{GCN}} & \multicolumn{1}{l|}{\textbf{GAT}} & \textbf{\begin{tabular}[c]{@{}l@{}}Graphs\\ age\end{tabular}} & \multicolumn{1}{l|}{\textbf{FdGars}} & \multicolumn{1}{l|}{\textbf{\begin{tabular}[c]{@{}l@{}}Player2\\ vec\end{tabular}}} & \multicolumn{1}{l|}{\textbf{\begin{tabular}[c]{@{}l@{}}Graph\\ Consis\end{tabular}}} & \textbf{GEM} & \multicolumn{1}{l|}{\textbf{\begin{tabular}[c]{@{}l@{}}CARE\\ -GNN\end{tabular}}} & \multicolumn{1}{l|}{\textbf{\begin{tabular}[c]{@{}l@{}}Graph\\ Smote\end{tabular}}} & \multicolumn{1}{l|}{\textbf{\begin{tabular}[c]{@{}l@{}}PC-\\ GNN\end{tabular}}} & \textbf{\begin{tabular}[c]{@{}l@{}}GAT\\ -COBO\end{tabular}} & \multicolumn{1}{l|}{\textbf{\scriptsize CSGNN$_{R}$}} & \textbf {\scriptsize CSGNN$_{C}$} & \textbf{\scriptsize CSGNN}                     \\ \cline{2-16} 
                                        & {\textbf{Macro AUC}} & \multicolumn{1}{l|}{0.9263}       & \multicolumn{1}{l|}{0.9243}       & 0.9159                                                        & \multicolumn{1}{l|}{0.7887}          & \multicolumn{1}{l|}{0.7467}                                                         & \multicolumn{1}{l|}{0.7985}                                                          & 0.8619       & \multicolumn{1}{l|}{0.9384}                                                       & \multicolumn{1}{l|}{0.8820}                                                         & \multicolumn{1}{l|}{0.9273}                                                     & 0.9391                                                       & \multicolumn{1}{l|}{0.9208}          & 0.9401          & \textbf{0.9539}                    \\ \cline{2-16} 
                                        & \textbf{Macro Recall}                     & \multicolumn{1}{l|}{0.8597}       & \multicolumn{1}{l|}{0.8585}       & 0.8564                                                        & \multicolumn{1}{l|}{0.7082}          & \multicolumn{1}{l|}{0.5618}                                                         & \multicolumn{1}{l|}{0.7288}                                                          & 0.8209       & \multicolumn{1}{l|}{0.8717}                                                       & \multicolumn{1}{l|}{0.8412}                                                         & \multicolumn{1}{l|}{0.8725}                                                     & 0.8867                                                       & \multicolumn{1}{l|}{0.8568}          & 0.8651          & \textbf{0.8878}                    \\ \cline{2-16} 
\multirow{-4}{*}{\textbf{Sichuan}}      & {\textbf{G-Mean}}    & \multicolumn{1}{l|}{0.8530}       & \multicolumn{1}{l|}{0.8529}       & 0.8447                                                        & \multicolumn{1}{l|}{0.6914}          & \multicolumn{1}{l|}{0.4181}                                                         & \multicolumn{1}{l|}{0.7187}                                                          & 0.8153       & \multicolumn{1}{l|}{0.8711}                                                       & \multicolumn{1}{l|}{0.8349}                                                         & \multicolumn{1}{l|}{0.8678}                                                     & 0.8829                                                       & \multicolumn{1}{l|}{0.8511}          & 0.8582          & \textbf{0.8864}                    \\ \hline
                                        & \textbf{Macro AUC}                        & \multicolumn{1}{l|}{0.8932}       & \multicolumn{1}{l|}{0.9102}       & 0.8928                                                        & \multicolumn{1}{l|}{0.6462}          & \multicolumn{1}{l|}{0.5227}                                                         & \multicolumn{1}{l|}{0.6211}                                                          & 0.6788       & \multicolumn{1}{l|}{0.9065}                                                       & \multicolumn{1}{l|}{0.8636}                                                         & \multicolumn{1}{l|}{0.9341}                                                     & 0.9109                                                       & \multicolumn{1}{l|}{0.9306}          & 0.7951          & \textbf{0.9698}                    \\ \cline{2-16} 
                                        & \textbf{Macro Recall}                     & \multicolumn{1}{l|}{0.5706}       & \multicolumn{1}{l|}{0.6152}       & 0.6715                                                        & \multicolumn{1}{l|}{0.4357}          & \multicolumn{1}{l|}{0.3206}                                                         & \multicolumn{1}{l|}{0.3311}                                                          & 0.3344       & \multicolumn{1}{l|}{0.7642}                                                       & \multicolumn{1}{l|}{0.7255}                                                         & \multicolumn{1}{l|}{0.7685}                                                     & 0.7823                                                       & \multicolumn{1}{l|}{0.7653}          & 0.6402          & \textbf{0.8317}                    \\ \cline{2-16} 
\multirow{-3}{*}{\textbf{BUPT}}         & \textbf{G-Mean}                           & \multicolumn{1}{l|}{0.4380}       & \multicolumn{1}{l|}{0.5267}       & 0.5823                                                        & \multicolumn{1}{l|}{0.3855}          & \multicolumn{1}{l|}{0.2502}                                                         & \multicolumn{1}{l|}{0.1602}                                                          & 0.0117       & \multicolumn{1}{l|}{0.7538}                                                       & \multicolumn{1}{l|}{0.7001}                                                         & \multicolumn{1}{l|}{0.7401}                                                     & 0.7658                                                       & \multicolumn{1}{l|}{0.7510}          & 0.5588          & \textbf{0.8262}                    \\ \hline
\end{tabular}}
\label{overallresult}
\end{table*}

\subsubsection{Comparison method}
We compared the proposed method with various GNN baselines in semi-supervised setting to confirm its efficacy. To be specific, we adopt the general GNN models: GCN \cite{kipf2016semi}, GAT \cite{velivckovic2017graph}, and GraphSAGE \cite{hamilton2017inductive}. The most sophisticated GNN-based fraud detectors: FdGars \cite{wang2019fdgars} Player2Vec \cite{zhang2019key}, GraphConsis \cite{liu2020alleviating}, and GEM \cite{liu2018heterogeneous}. The most advanced imbalance graph learners: PC-GNN \cite{liu2021pick}, GraphSmote \cite{zhao2021graphsmote}, CARE-GNN \cite{dou2020enhancing}, and GAT-COBO \cite{hu2022gatcobo}. Since our graphs are homogeneous, all the above methods are computed in homogeneous way.

\begin{itemize}

\item GCN: A GNN that uses spectral graph convolution to aggregate neighbor information.

\item GAT: A GNN that uses an attention mechanism to aggregate information about neighbor nodes.

\item GraphSAGE: An inductive GNN with a fixed number of sampled neighbors.

\item FdGars: A GCN-based social opinion fraud detection system.

\item Player2Vec: A GNN with heterogeneous information networks and Meta Path.

\item GraphConsis: A heterogeneous GNN for resolving inconsistencies in graphs.

\item GEM: A GNN for heterogeneous graph fraud detection based on the attention mechanism.

\item CARE-GNN: A heterogeneous GNN based on reinforcement learning neighbor sampling.

\item GAT-COBO: A cost-sensitive GNN based on GNN and Adaboost.

\item GraphSmote: An imbalanced GNN based on the Smote sampling algorithm.

\item PC-GNN: an unbalanced GNN with node and edge resampling.

\item CSGNN: Our proposed method.

\item CSGNN$_{R}$: CSGNN with the RL-based neighbor sampling module removed.

\item CSGNN$_{C}$: CSGNN with the cost-sensitive learning module removed.

\end{itemize}

\subsubsection{ Evaluation metrics and experimental setup}\label{experimental setup}
For the assessment of imbalance problems, evaluation metrics and datasets are critical. In such problems, we usually focus more on positive instances (e.g., fraudsters). To avoid bias, we adopt three widely used metrics to measure the performance of all compared methods: Macro AUC, Macro recall, and G-mean.

In our experiments, we choose the training samples randomly and keep the ratio of positive and negative samples in the training set the same as the entire datasets. In order to optimize the parameters, we utilize the Adam optimizer. And other hyper-parameters are set up as follows:
For the Sichuan dataset, we set hid embedding size(256), learning rate(0.01), model layers(3), max epoch(15). For the BUPT dataset, we set hid embedding size(64), learning rate(0.01), model layers(5), max epoch(1). We implement the proposed method by Pytorch; all models are running on python3.7.10, 1 GeForce RTX 3090 GPU, 64GB RAM, 16 cores Intel(R) Xeon(R) Gold 5218 CPU @2.30GHz Linux Server.

\begin{figure*}[htbp]

\centering
\subfigure[Imbalance Rate]{
\begin{minipage}{5cm}
\label{G-mean sichuan}
\includegraphics[width=5cm]{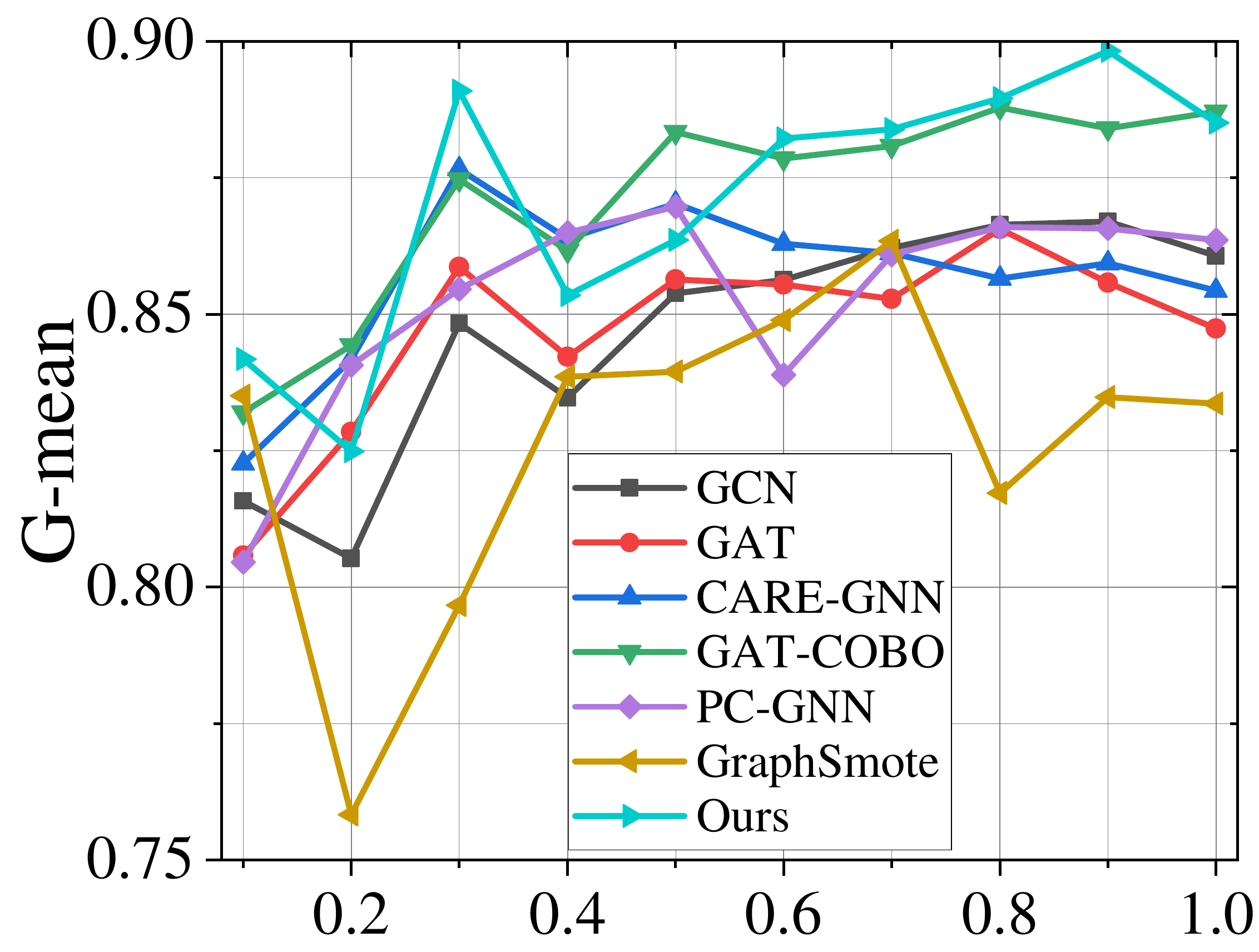}
\end{minipage}
}
\hspace{0mm}
\subfigure[Imbalance Rate]{
\begin{minipage}{5cm}
\centering      
\label{AUC sichuan}
\includegraphics[width=5cm]{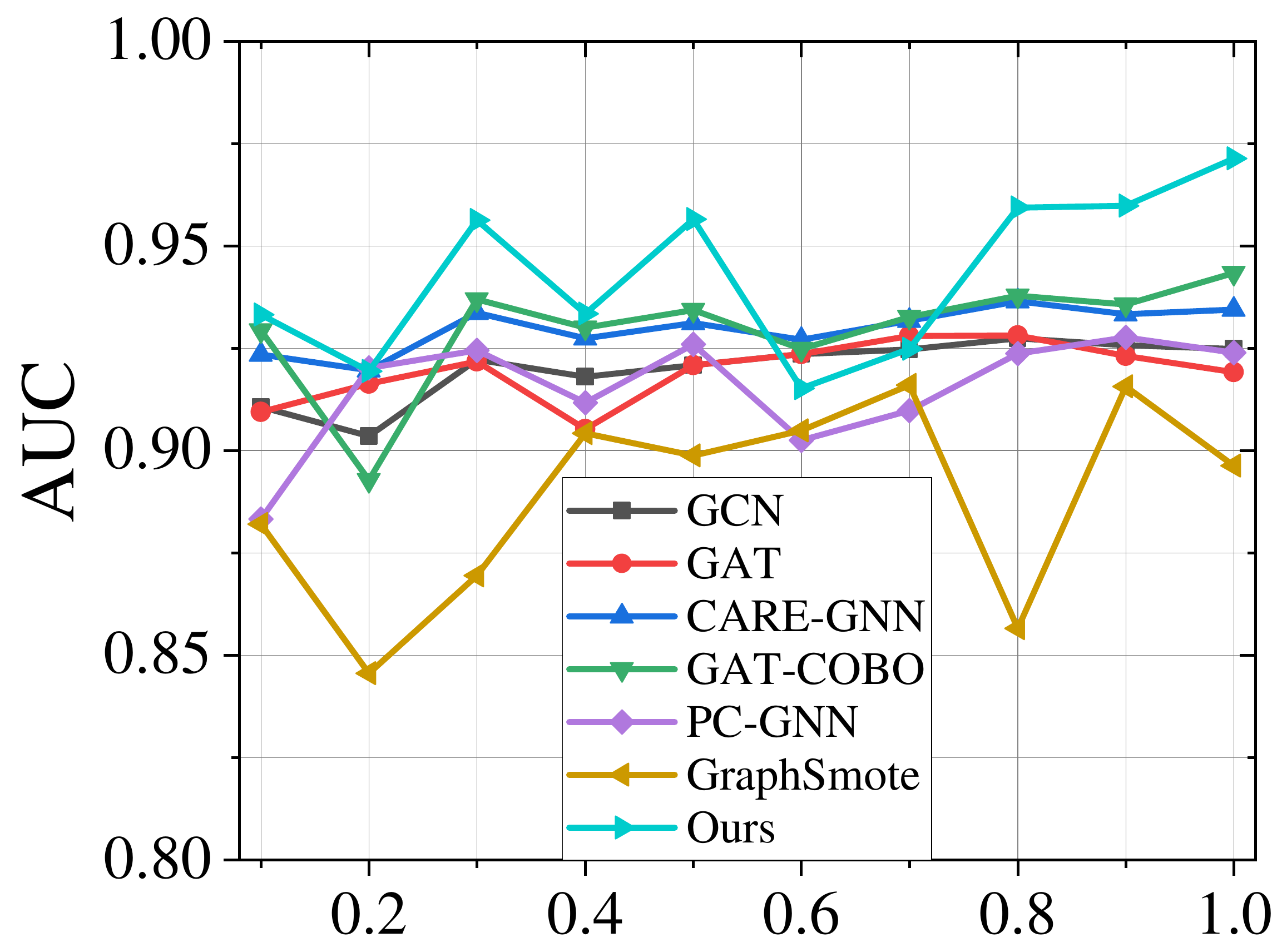}
\end{minipage}
}
\hspace{0mm}
\subfigure[Imbalance Rate]{
\begin{minipage}{5cm}
\label{AUC sichuan}
\includegraphics[width=5cm]{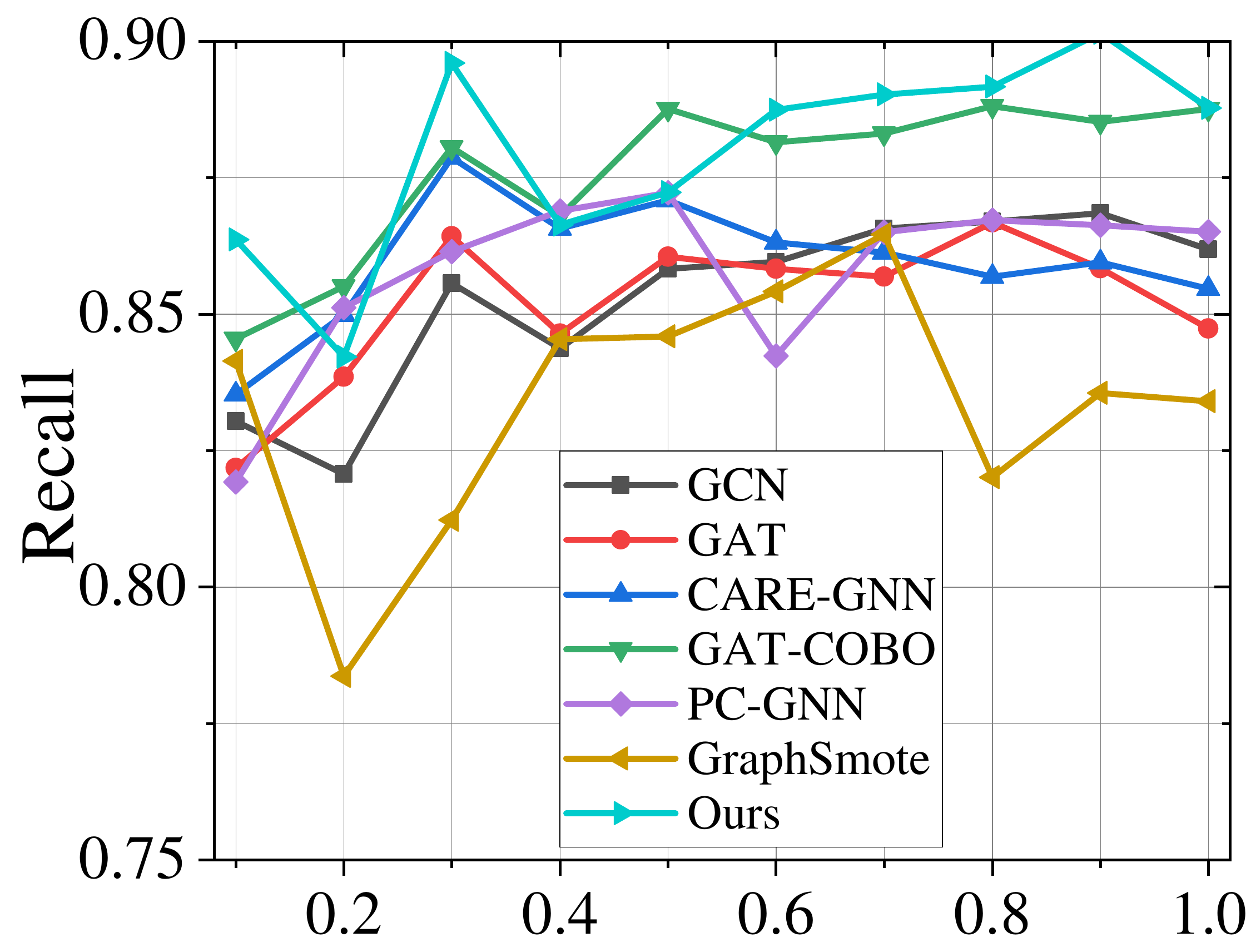}
\end{minipage}
}

\hspace{0mm}
\subfigure[Imbalance Rate]{
\begin{minipage}{5cm}
\label{G-mean bupt}
\includegraphics[width=5cm]{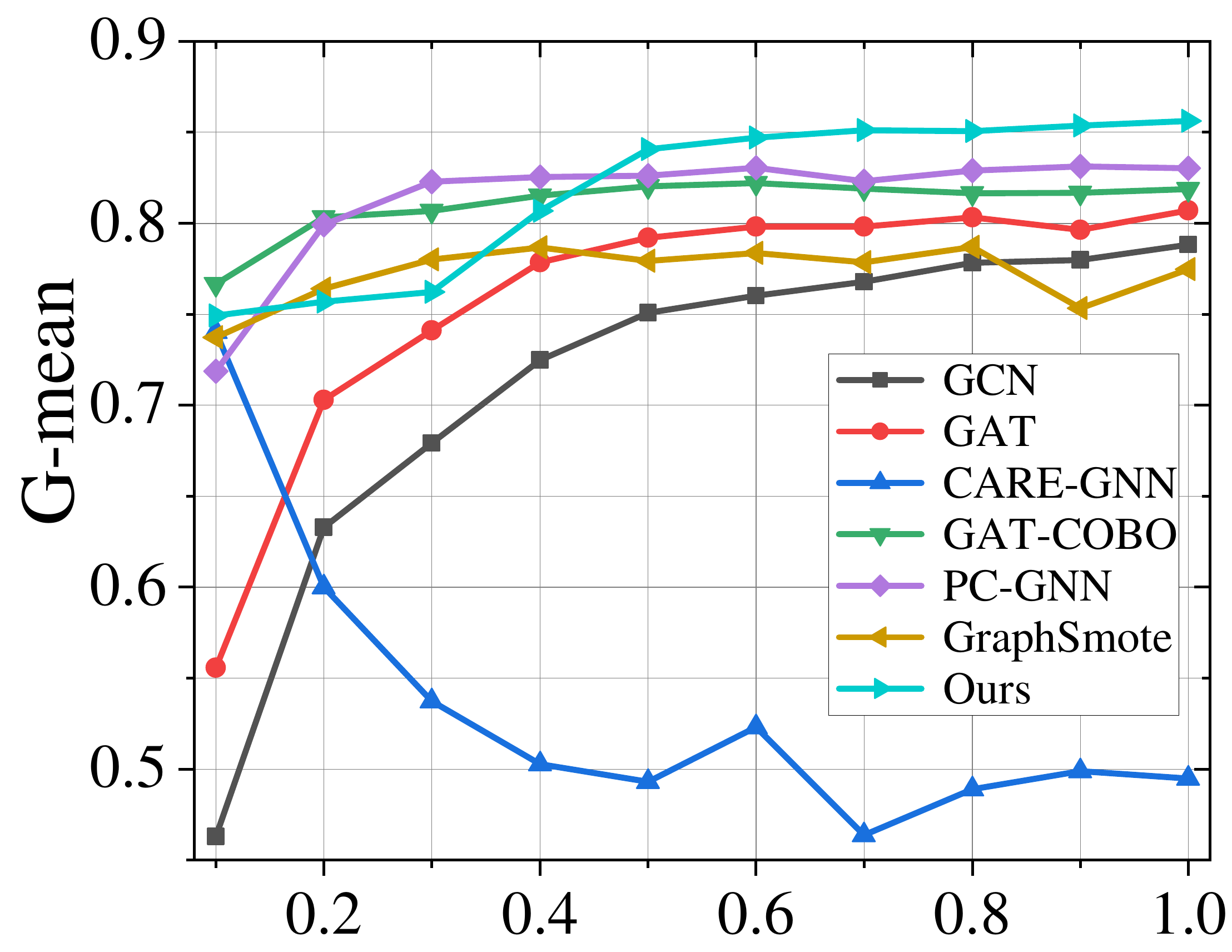}
\end{minipage}
}
\hspace{0mm}
\subfigure[Imbalance Rate]{
\begin{minipage}{5cm}
\label{AUC bupt}
\includegraphics[width=5cm]{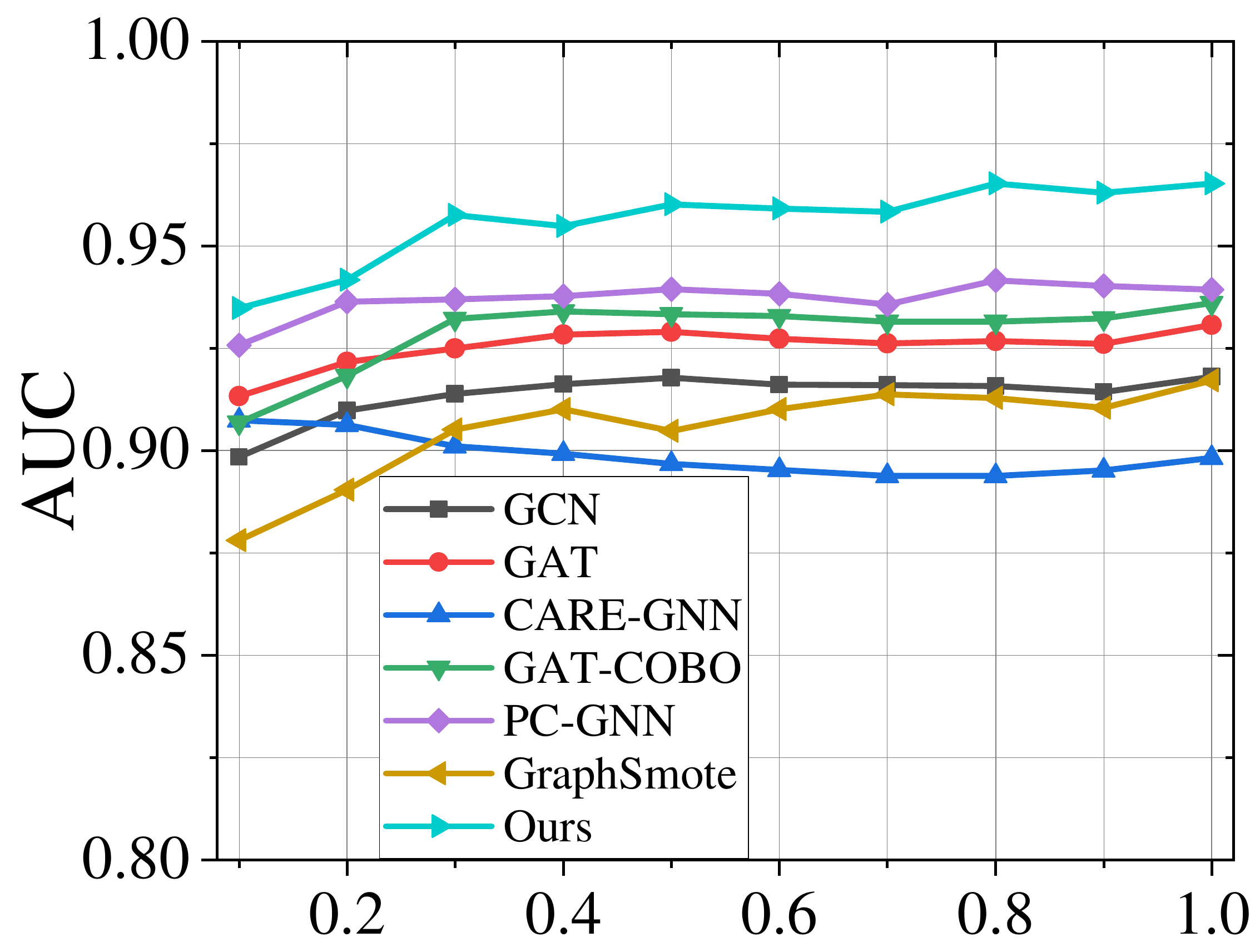}
\end{minipage}
}
\hspace{0mm}
\subfigure[Imbalance Rate]{
\begin{minipage}{5cm}
\label{AUC sichuan}
\includegraphics[width=5cm]{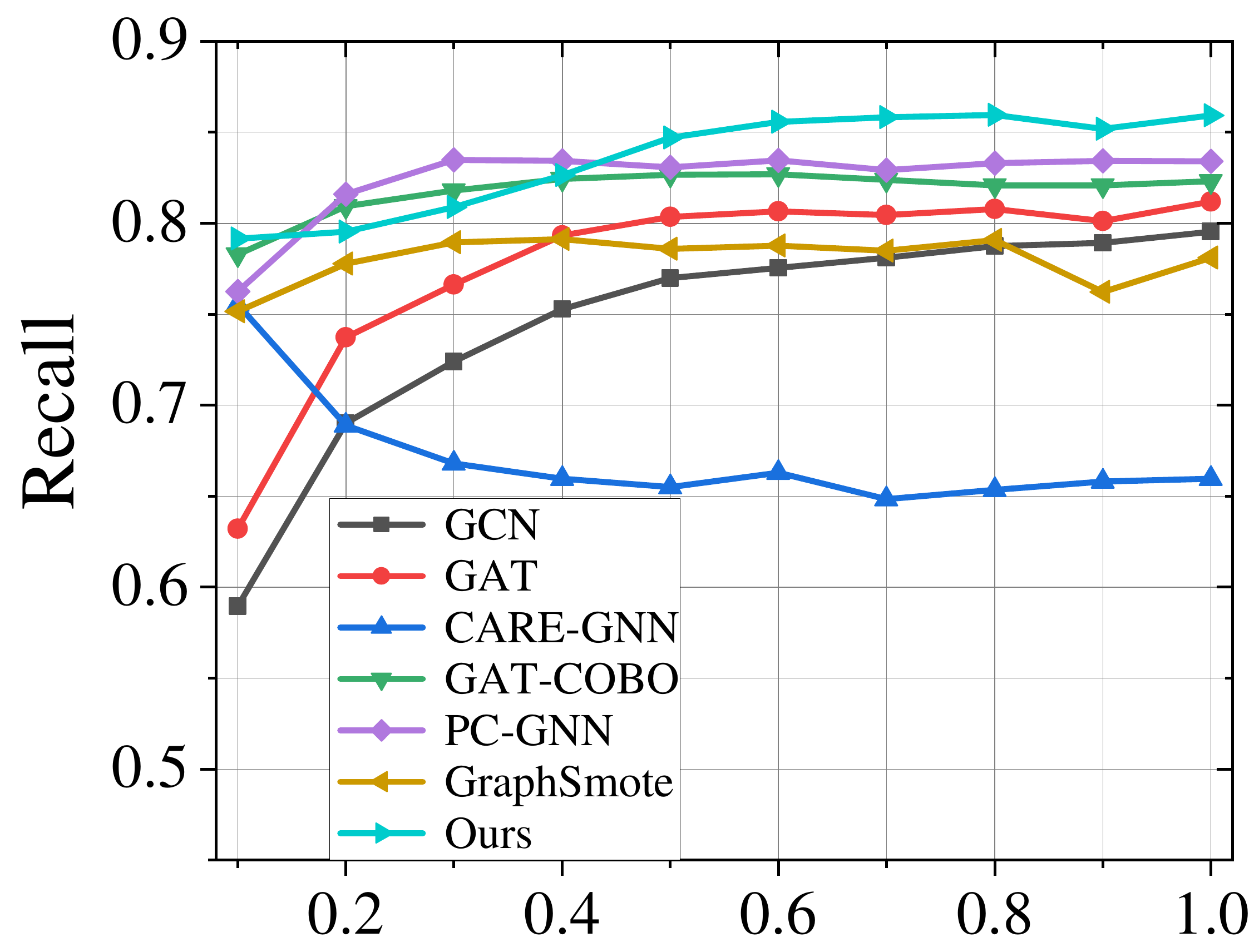}
\end{minipage}
}
\caption{Performance comparison of baseline methods with different IR. The top row is on the Sichuan dataset, and the bottom is on BUPT. }
\label {Performance comparison}
\end{figure*}

\subsection{Overall Evaluation(RQ1)}
\label{Overall Evaluation}

In our experiments, we set the ratio of training set, validation set and test set to 0.2:0.2:0.6. Table \ref{table3} demonstrates the scores of the baseline approaches and the proposed model on the three evaluation metrics. It can be observed that our proposed method CSGNN scores better than the comparison method on all metrics on two real-world telecom fraud detection datasets. In addition, we observe the following phenomena.

First, we observe that the imbalanced graph learner outperforms the general GNN and GNN-based fraud detectors. To be more specific, CARE-GNN, PC-GNN, and GAT-COBO have higher scores on all metrics than the general GNN and GNN-based fraud detectors. GraphSmote also outperforms the general GNN and GNN-based fraud detectors greatly on BUPT, although marginally underperforming the general GNN on the Sichuan dataset. The intuitive explanation for this phenomenon is that the minority class samples in the graph can be better learned by the imbalanced graph learner, allowing the model to remain unbiased. Imbalance is a well-known issues in fraud detection, and it is also the key factor in degrading model performance. This implies that designing particularly for the graph imbalance issue is crucial for the fraud detection problem in order to enhance model performance.

Moreover, general GNNs such as GCN, GAT, and GraphSAGE perform better than GNN-based fraud detectors such as FdGars, Player2Vec, GraphConsis, and GEM on two telecom fraud datasets. The reasons for this phenomenon can be broadly classified into the following aspects: First, general GNNs with basic, straightforward design principles and strong universalities, such as GCN, GAT, and GraphSAGE, are designed to execute deep learning for general graphs. As a result, they have great performance in the majority of application cases. Second, GNN models that are customized for domain problems (FdGars, Player2Vec, GraphConsis, GEM) have large variations in performance on different graphs. These GNN models are designed for different domain problems (e.g., social, financial, insurance, gaming, etc.) and thus have domain characteristics. When migrating them to other domains, the performance is often not guaranteed. For example, models designed for financial fraud (e.g., GEM) and social fraud (e.g., FdGars) both perform so-so on the telecom fraud problem and vary widely, even though they are both fraud detectors. Third, GNN performance is closely related to graph type (e.g., homogeneous and heterogeneous graphs). The dataset of telecom fraud detection used in this paper is a homogeneous graph, while FdGars, Player2Vec, GraphConsis, and GEM are designed for heterogeneous graphs, which may lead to their performance degradation. This further indicates that the fraud detection methods on graphs have strong scenario dependencies.

Furthermore, our approach outperforms general GNNs, GNN-based fraud detectors, and the imbalanced graph learners. This further demonstrates the benefit of the proposed method in dealing with imbalanced graphs. Despite the fact that our technique is an imbalanced graph learner as well, we adopt two strategies to enhance the model's capacity to handle imbalanced graphs. On the one hand, we apply reinforcement learning for node neighbor sampling. On the other hand, we introduce cost-sensitive learning into GNN to make the model more focused on the minority classes. This will be a reference for the subsequent design of other GNNs.

\begin{table*}[htbp]
\centering
\caption{Macro recall score of each class in the ablation study.}
\renewcommand{\arraystretch}{1.9}
\setlength{\tabcolsep}{2.8mm}{
\begin{tabular}{|c|cc|ccc|}

\hline
Dataset & \multicolumn{2}{c|}{Sichuan}         & \multicolumn{3}{c|}{BUPT}                                          \\ \hline
Metric & \multicolumn{1}{c|}{Recall-Benign}   & Recall-Fraud    & \multicolumn{1}{c|}{Recall-Benign}   & \multicolumn{1}{c|}{Recall-Fraud}    & Recall-Courier  \\ \hline
CSGNN$_{R}$ & \multicolumn{1}{c|}{0.9547} & 0.7589 & \multicolumn{1}{c|}{0.8531} & \multicolumn{1}{c|}{0.8764} & 0.5665 \\ \hline
CSGNN$_{C}$ & \multicolumn{1}{c|}{\textbf{0.9740}} & 0.7562          & \multicolumn{1}{c|}{\textbf{0.9603}} & \multicolumn{1}{c|}{0.7012}          & 0.2592          \\ \hline
CSGNN  & \multicolumn{1}{c|}{0.9685}          & \textbf{0.7748} & \multicolumn{1}{c|}{0.7208}          & \multicolumn{1}{c|}{\textbf{0.9539}} & \textbf{0.8203} \\ \hline
\end{tabular}}
\label{ablation}
\end{table*}

\subsection{Influence of Imbalance Ratio(RQ2)}
The imbalance graph learner should have superior performance at different imbalance rates. To further test the performance of CSGNN on imbalanced graphs, we validated it by setting up a set of experiments. Specifically, we take a value for IR in the range [0,1] at every interval of 0.1. In this way, we obtain 10 different groups of IR values. Then, in order to create 10 new datasets that match the IR values, samples from various classes in the two datasets are sequentially randomly sampled according to these 10 sets of IRs. Among all of the baseline approaches, we selected six that performed best in \ref{Overall Evaluation}, which include two general-purpose GNNs (GCN and GAT) and four imbalanced graph learners (CARE-GNN, GAT-COBO, PC-GNN, and GraphSmote), and evaluated them alongside the approach proposed in this paper. We recorded the scores of the above methods on three metrics, G-Mean, AUC, and Recall, and plotted them as the curves shown in Fig.\ref {Performance comparison}.

It can be observed that in most cases, our proposed method in this paper outperforms the comparison method for different imbalance rates, both on the Sichuan dataset and the BUPT dataset. When the IR decreases, the scores of almost all models start to decrease, but the proposed method still outperforms the comparison methods. In addition, the general GNN models (GCN, GAT) also have good performance, but when the IR decreases, the performance of these models decreases faster than the proposed method. We can also observe that there are small fluctuations in the performance of all models under different IR and that the fluctuation trend is similar. This is due to the different datasets sampled by different IR. Nevertheless, the proposed CSGNN continues to perform better even when using different samples of data. This illustrates the significance of the special design of the proposed method for imbalanced graphs. In particular, the neighbor sampling strategy based on reinforcement learning balances the data distribution, and the cost-sensitive matrix adjusts GNN embedding, making the model more robust in the face of unbalanced data.

We also observe that most of the imbalanced graph learners generally perform better than the general GNN for different IR values, which indicates the essential role of cost sensitive design in imbalanced graph. In particular, PC-GNN and GAT-COBO show excellent performance. The former uses oversampling and undersampling of nodes and connected edges to cope with unbalanced graphs, while the latter combines cost-sensitive boosting methods with GNNs to overcome model bias in graph machine learning. In this paper, cost-sensitive learning is also used to solve the imbalance problem in graphs. This illustrates the effectiveness of imbalance learning methods in various types of data, even unstructured data like graphs. This is an important research direction for future graph machine learning.

\subsection{Ablation Study(RQ3)}
As mentioned in Section \ref{Method}, the proposed CSGNN contains two key modules: the neighbor sampler and the cost-sensitive learner. To clarify the contributions of two modules in CSGNN, we verify their effectiveness by removing the neighbor sampler (CSGNN$_{R}$) and the cost-sensitive learner (CSGNN$_{C}$), respectively. The experimental results are shown in the Ablation column in Table \ref{overallresult}. It can be seen that CSGNN scores the highest on all three metrics compared to CSGNN$_{R}$ and CSGNN$_{C}$. This illustrates that each part of our model is crucial to improve the final performance. Besides, we also observe other interesting phenomena.

On the Sichuan dataset, CSGNN$_{R}$ and CSGNN$_{C}$ scored roughly $1\%$–$3\%$ lower than those of CSGNN. This finding suggests that, while there is a disparity between the performances of the two methods, it is not statistically significant. However, the situation is very different on BUPT dataset. On the three metrics, the score of CSGNN$_{R}$ is $3\%$–7$\%$ lower than that of CSGNN, while the score of CSGNN$_{C}$ was up to $17\%$–27$\%$ lower than that of CSGNN. It is an intriguing issue to investigate the cause of the huge difference in the performance of CSGNN on the two telecom fraud datasets.

To explore the reason, we present Recall-scores for each class in the ablation study experiment in Table \ref{ablation}. As can be seen, CSGNN$_{C}$ scores higher on Recall-Benign but lower on Recall-Fraud and Recall-Courier. In particular, in the BUPT dataset, CSGNN$_{C}$scores 57$\%$ lower than CSGNN on Recall-Courier and 25$\%$ lower on Recall-Fraud. This indicates that CSGNN$_{C}$ has a poor prediction for the minority classes of Fraud and Courier. Similarly, CSGNN$_{R}$ is much inferior to CSGNN in predicting the minority classes of Fraud and Courier. In addition, the performance of CSGNN$_{R}$ and CSGNN$_{C}$ on BUPT decreases sharply compared to the Sichuan dataset. The reason is that the $IR$ values of the two datasets are very different from each other. More precisely, IR$_{Sichuan}$(0.4735)$>$IR$_{BUPT}$(0.0809). Due to the extreme imbalance of BUPT, the performance of CSGNN$_{R}$ and CSGNN$_{C}$ without the imbalance module cannot be guaranteed. In contrast, CSGNN is able to handle the imbalance in the graph easily. The above phenomenon further illustrates the important functions served by the neighbor sampling and cost-sensitive modules in CSGNN. Moreover, the worse the dataset balance, the more significant the CSGNN performance improvement.

\subsection{Sensitive Analysis(RQ4)}

\begin{figure}[ht]
\centering

\subfigure[Train Size]{
\label{Train}
\includegraphics[width=4cm]{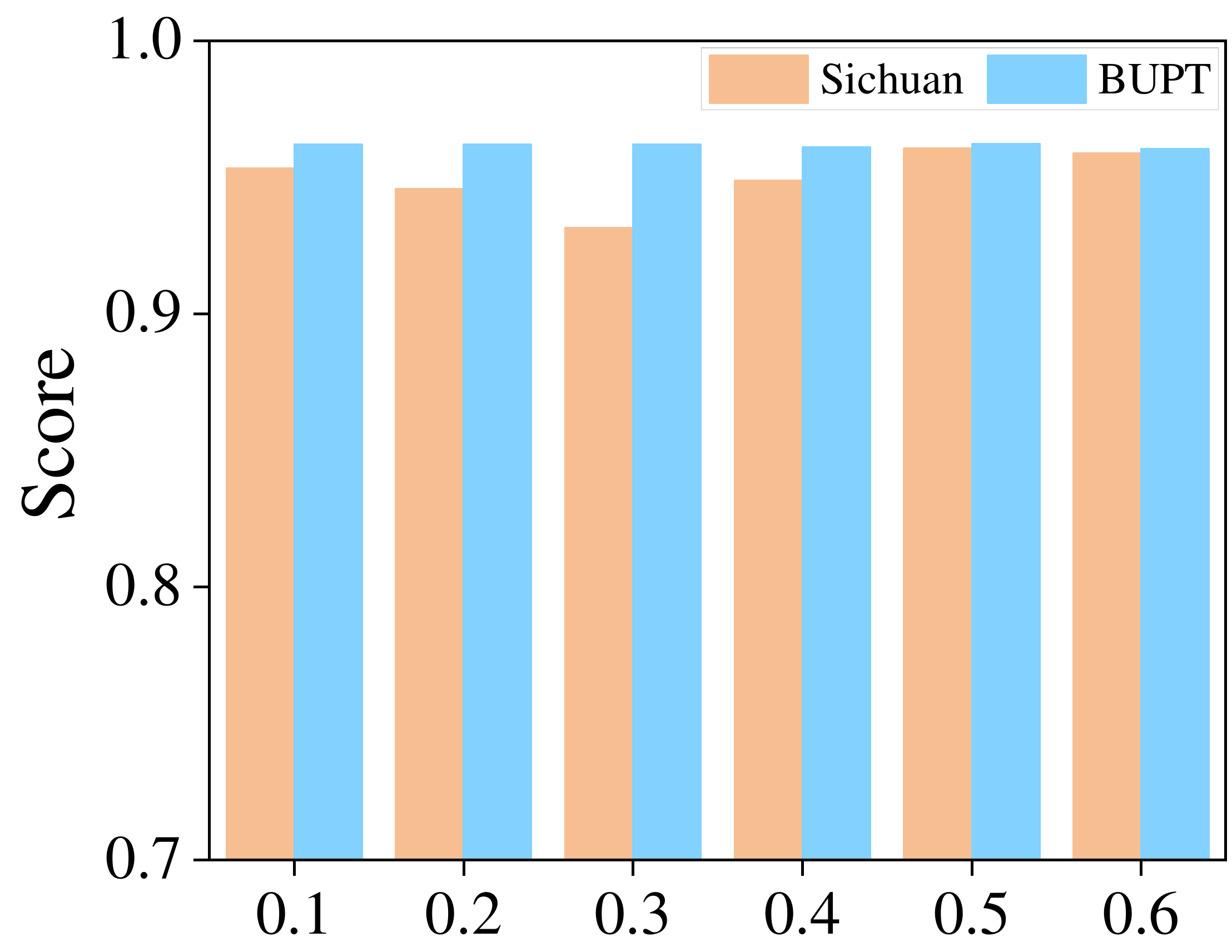}
}
\hspace{0mm}
\subfigure[$\beta$]{
\label{beta}
\includegraphics[width=4cm]{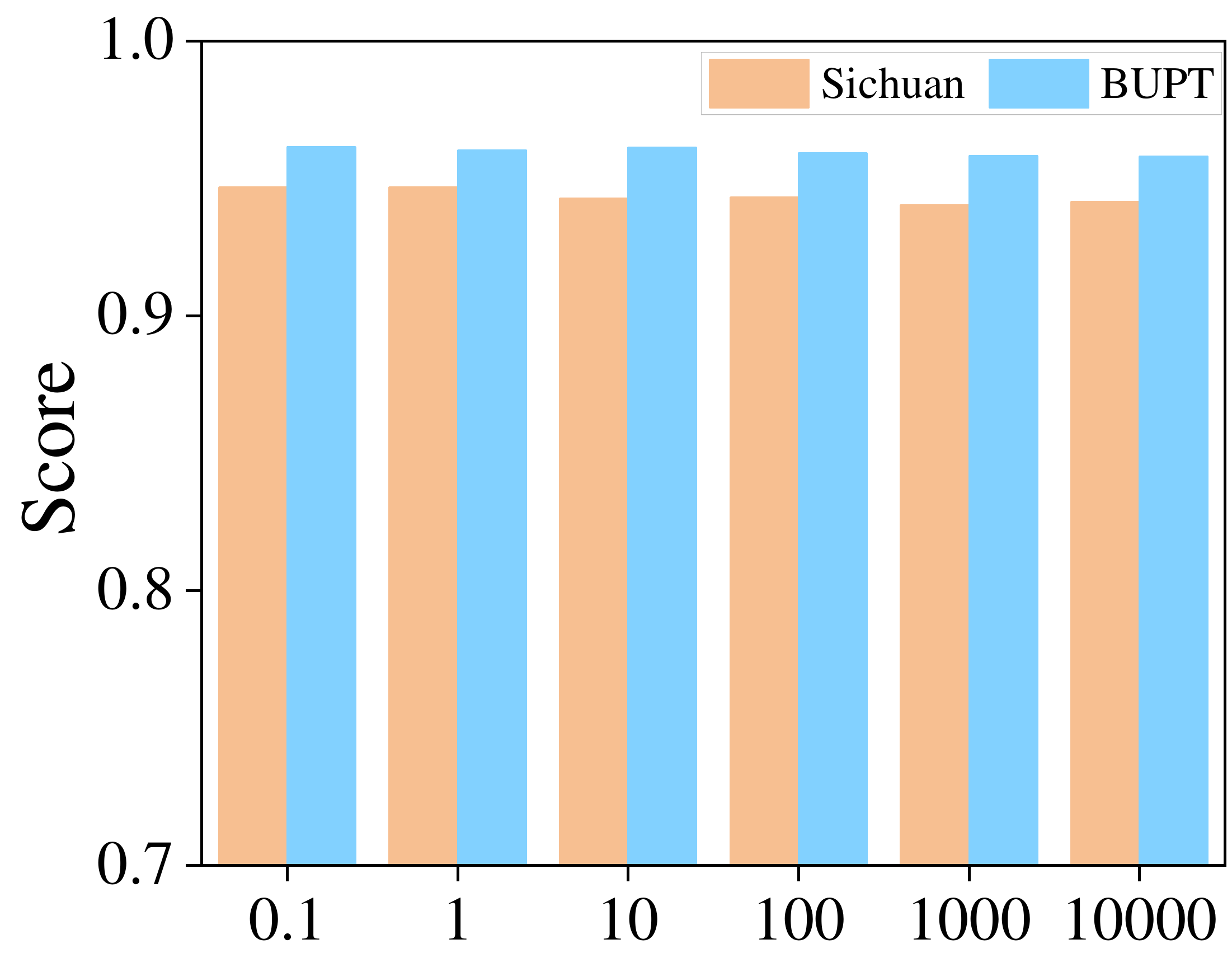}
}
\hspace{0mm}
\subfigure[Layer]{
\label{Layer}
\includegraphics[width=4cm]{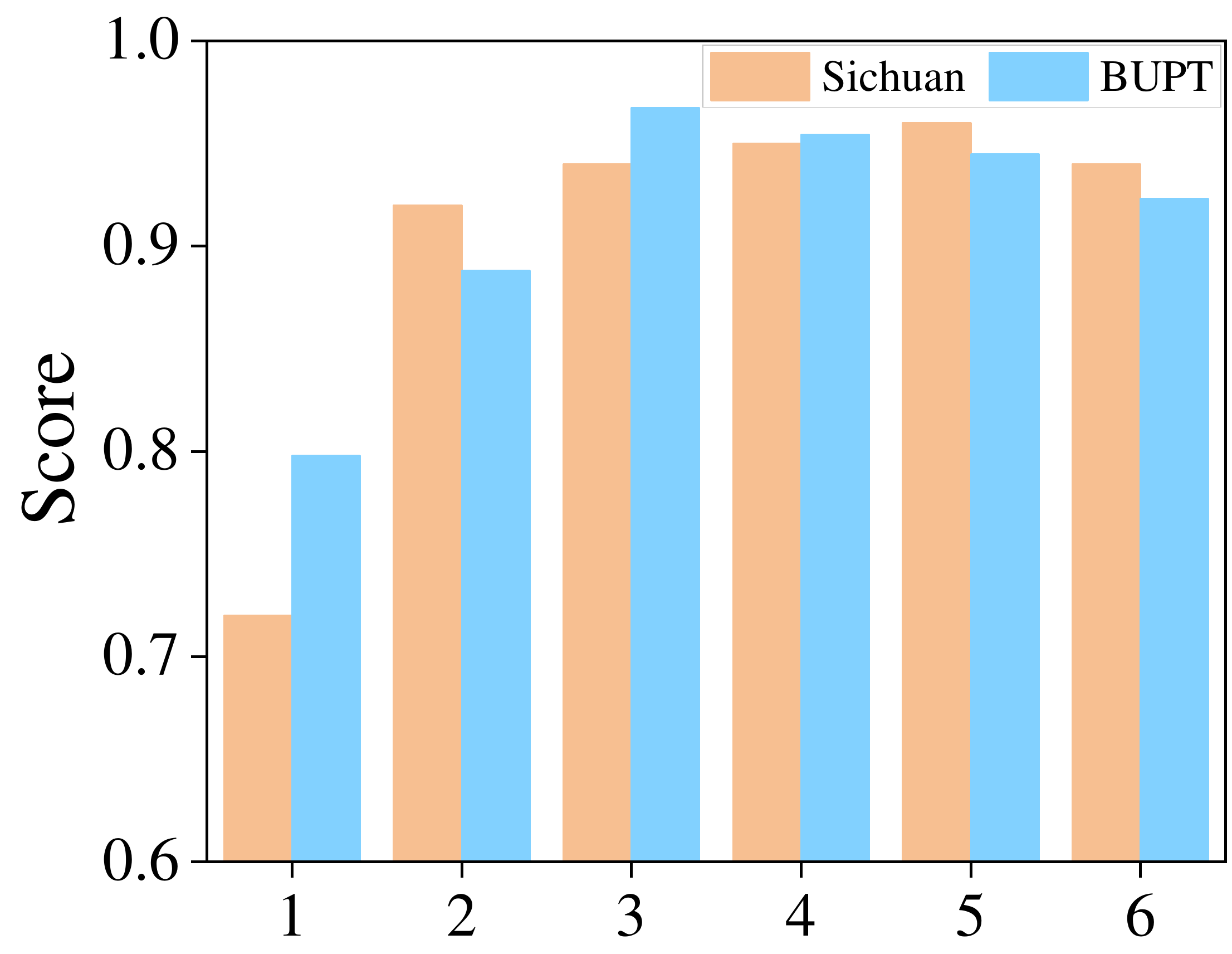}
}
\hspace{0mm}
\subfigure[hid dimension]{
\label{hid}
\includegraphics[width=4cm]{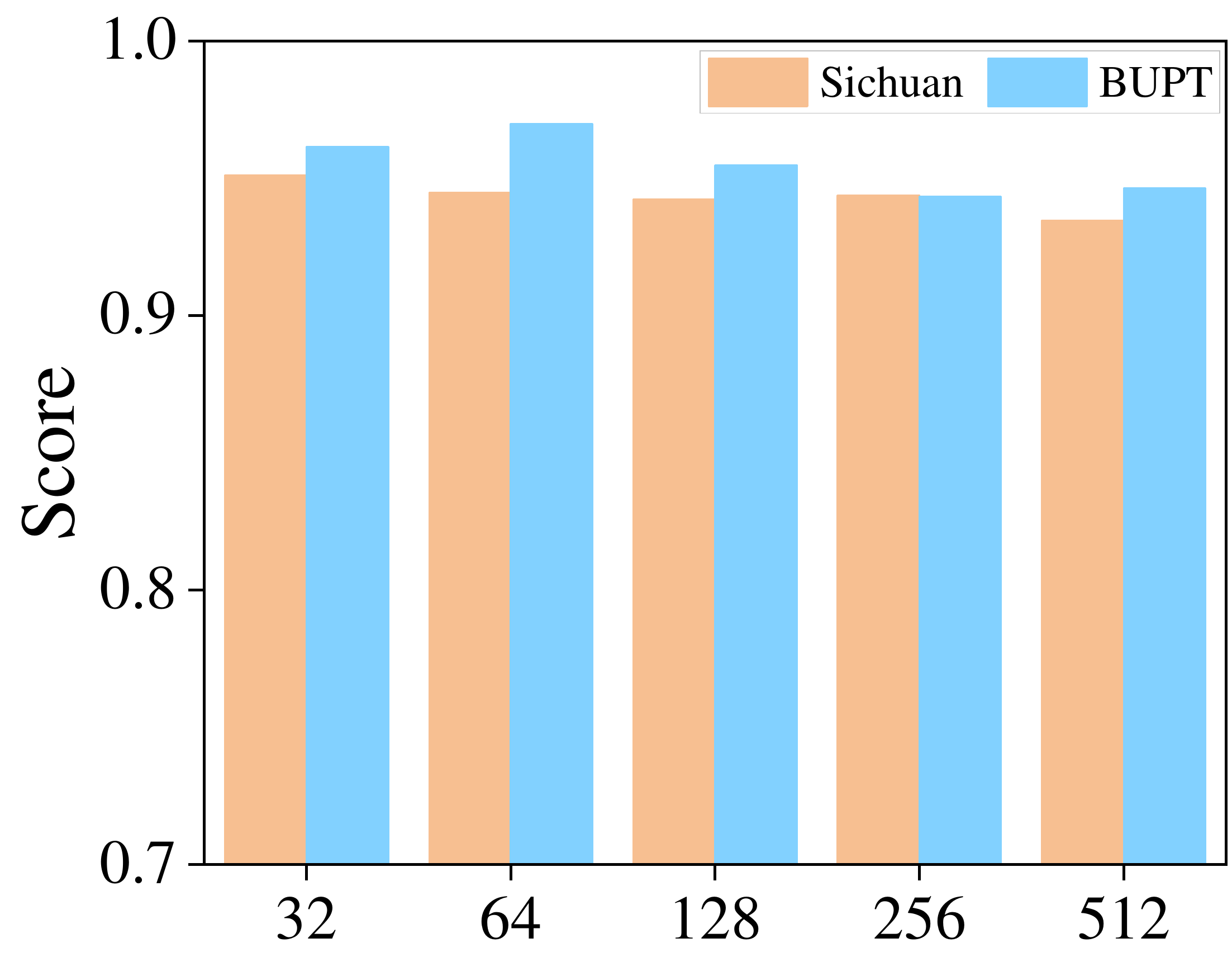}
}
\caption{Hyperparameters Sensitivity}
\label {Sensitivity}
\end{figure}

Fig.\ref{Sensitivity} shows the performance of CSGNN about the four hyperparameters on Sichuan and BUPT datasets. From Fig. \ref{Train}, we observe that increasing the training set proportion hardly improves the performance of CSGNN. CSGNN achieves better performance with a 10$\%$ training set proportion, which shows that a small amount of training data is sufficient for CSGNN. This allows the model to reduce its dependence on data as well as save computational costs.  Fig.\ref{beta} shows the performance of CSGNN regarding the hyperparameter $\beta$ in Section \ref{Cost-sensitive learning module}. It can be seen that the values of $\beta$ have little effect on the model performance.  Fig.\ref{Layer} shows the influence of the number of layers of the neural network. We can observe that the performance gradually improves when layer=1$\sim$3. CSGNN performs best when the number of layers is around 3. And when the number of layers is too high, the performance starts to decrease. This indicates that  about 3 layers can make CSGNN perform better, and too many layers will not improve performance and increase the computational cost.  Fig.\ref{hid} shows the effect of the number of neurons in the hidden layer. It can be observed that when hid unit=32$\sim$64, the model can achieve better performance. More neurons in the hidden layer will not improve the performance.

 \section{Conclusion}
In this paper, we study fraud detection in mobile social networks. To address the graph imbalance problem, we introduced cost-sensitive learning into graph neural networks and proposed a cost-sensitive graph neural network framework CSGNN. First, we designed a reinforcement learning strategy for node neighborhood sampling, which is used to initially mitigate the data imbalance in node neighborhoods. Then we designed a cost matrix learning method and combined it with GNN, which can automate the learning to optimal node embedding. Finally, cost-sensitive embedding is used for downstream fraud detection task. As a result of solving the graph imbalance problem, our model achieves significant performance improvements compared to several of the most advanced baseline methods. It is believed that cost-sensitive learning is a very promising solution to the graph imbalance problem although there is still little work of this kind. We look forward to more related research in the future.

\section*{Author Contributions}
Xinxin Hu: Conceptualization, formal analysis, methodology, software, writing-original draft. Haotian Chen: Formal analysis, methodology, software. Hongchang Chen: Funding acquisition, supervision. Shuxin Liu: Conceptualization, validation. Xing Li: Writing- review and editing. Shibo Zhang: Writing-original draft. Yahui Wang: Investigation, Software. Xiangyang Xue: Conceptualization, supervision.

\section*{Acknowledgements}
We would like to express our gratitude to Siqi Yang for his valuable suggestions. We would like to thank anonymous reviewers for their valuable comments.

\section*{Conflict of Interest}
The authors have declared no conflict of interest.

\section*{Data availability Statement}
The datasets generated or analyzed during this study are available at \textcolor{blue}{\url{https://github.com/xxhu94/CSGNN}}.

\small
\bibliographystyle{IEEEtran}
\bibliography{IEEEabrv,sample-base}

\begin{thebibliography}{10}
\providecommand{\url}[1]{#1}
\csname url@samestyle\endcsname
\providecommand{\newblock}{\relax}
\providecommand{\bibinfo}[2]{#2}
\providecommand{\BIBentrySTDinterwordspacing}{\spaceskip=0pt\relax}
\providecommand{\BIBentryALTinterwordstretchfactor}{4}
\providecommand{\BIBentryALTinterwordspacing}{\spaceskip=\fontdimen2\font plus
\BIBentryALTinterwordstretchfactor\fontdimen3\font minus
  \fontdimen4\font\relax}
\providecommand{\BIBforeignlanguage}[2]{{%
\expandafter\ifx\csname l@#1\endcsname\relax
\typeout{** WARNING: IEEEtran.bst: No hyphenation pattern has been}%
\typeout{** loaded for the language `#1'. Using the pattern for}%
\typeout{** the default language instead.}%
\else
\language=\csname l@#1\endcsname
\fi
#2}}
\providecommand{\BIBdecl}{\relax}
\BIBdecl

\bibitem{yang2019mining}
Y.~Yang, Y.~Xu, Y.~Sun, Y.~Dong, F.~Wu, and Y.~Zhuang, ``Mining fraudsters and
  fraudulent strategies in large-scale mobile social networks,'' \emph{IEEE
  Transactions on Knowledge and Data Engineering}, vol.~33, no.~1, pp.
  169--179, 2019.

\bibitem{zheng2018generative}
Y.-J. Zheng, X.-H. Zhou, W.-G. Sheng, Y.~Xue, and S.-Y. Chen, ``Generative
  adversarial network based telecom fraud detection at the receiving bank,''
  \emph{Neural Networks}, vol. 102, pp. 78--86, 2018.

\bibitem{caict2020fraud}
``Research report on telecommunication network fraud management under the new
  situation,'' China Academy of Information and Communications Technology,
  Tech. Rep., 2020.

\bibitem{truecaller2022spam}
``Truecaller insights 2022 u.s. spam \& scam report,'' Truecaller Insights,
  Tech. Rep., 2022.

\bibitem{whoscall2022latest}
``Latest analysis on global scam calls and messages from global and local
  perspectives,'' Whoscall, Tech. Rep., 2022.

\bibitem{hooi2016fraudar}
B.~Hooi, H.~A. Song, A.~Beutel, N.~Shah, K.~Shin, and C.~Faloutsos, ``Fraudar:
  Bounding graph fraud in the face of camouflage,'' in \emph{Proceedings of the
  22nd ACM SIGKDD international conference on knowledge discovery and data
  mining}, 2016, pp. 895--904.

\bibitem{tseng2015fraudetector}
V.~S. Tseng, J.-C. Ying, C.-W. Huang, Y.~Kao, and K.-T. Chen, ``Fraudetector: A
  graph-mining-based framework for fraudulent phone call detection,'' in
  \emph{Proceedings of the 21th ACM SIGKDD International Conference on
  Knowledge Discovery and Data Mining}, 2015, pp. 2157--2166.

\bibitem{huang2022graph}
Z.~Huang, Y.~Tang, and Y.~Chen, ``A graph neural network-based node
  classification model on class-imbalanced graph data,'' \emph{Knowledge-Based
  Systems}, vol. 244, p. 108538, 2022.

\bibitem{wu2021graphmixup}
L.~Wu, H.~Lin, Z.~Gao, C.~Tan, S.~Li \emph{et~al.}, ``Graphmixup: Improving
  class-imbalanced node classification on graphs by self-supervised context
  prediction,'' \emph{arXiv preprint arXiv:2106.11133}, 2021.

\bibitem{zhao2021graphsmote}
T.~Zhao, X.~Zhang, and S.~Wang, ``Graphsmote: Imbalanced node classification on
  graphs with graph neural networks,'' in \emph{Proceedings of the 14th ACM
  international conference on web search and data mining}, 2021, pp. 833--841.

\bibitem{liu2021pick}
Y.~Liu, X.~Ao, Z.~Qin, J.~Chi, J.~Feng, H.~Yang, and Q.~He, ``Pick and choose:
  a gnn-based imbalanced learning approach for fraud detection,'' in
  \emph{Proceedings of the Web Conference 2021}, 2021, pp. 3168--3177.

\bibitem{khan2017cost}
S.~H. Khan, M.~Hayat, M.~Bennamoun, F.~A. Sohel, and R.~Togneri,
  ``Cost-sensitive learning of deep feature representations from imbalanced
  data,'' \emph{IEEE transactions on neural networks and learning systems},
  vol.~29, no.~8, pp. 3573--3587, 2017.

\bibitem{2021Network}
Z.~Wang, X.~Ye, C.~Wang, J.~Cui, and P.~S. Yu, ``Network embedding with
  completely-imbalanced labels,'' \emph{Institute of Electrical and Electronics
  Engineers (IEEE)}, no.~11, 2021.

\bibitem{santos2022facs}
F.~Santos, J.~Ye, F.~Masrour, P.-N. Tan, and A.-H. Esfahanian, ``Facs-gcn:
  Fairness-aware cost-sensitive boosting of graph convolutional networks,'' in
  \emph{2022 International Joint Conference on Neural Networks (IJCNN)}.\hskip
  1em plus 0.5em minus 0.4em\relax IEEE, 2022, pp. 1--8.

\bibitem{duan2022dual}
Y.~Duan, X.~Liu, A.~Jatowt, H.-t. Yu, S.~Lynden, K.-S. Kim, and A.~Matono,
  ``Dual cost-sensitive graph convolutional network,'' in \emph{2022
  International Joint Conference on Neural Networks (IJCNN)}.\hskip 1em plus
  0.5em minus 0.4em\relax IEEE, 2022, pp. 1--8.

\bibitem{wang2022imbalanced}
Y.~Wang, Y.~Zhao, N.~Shah, and T.~Derr, ``Imbalanced graph classification via
  graph-of-graph neural networks,'' in \emph{Proceedings of the 31st ACM
  International Conference on Information \& Knowledge Management}, 2022, pp.
  2067--2076.

\bibitem{li2022graph}
X.~Li, L.~Wen, Y.~Deng, F.~Feng, X.~Hu, L.~Wang, and Z.~Fan, ``Graph neural
  network with curriculum learning for imbalanced node classification,''
  \emph{arXiv preprint arXiv:2202.02529}, 2022.

\bibitem{wang2021distance}
Y.~Wang, C.~Aggarwal, and T.~Derr, ``Distance-wise prototypical graph neural
  network in node imbalance classification,'' \emph{arXiv preprint
  arXiv:2110.12035}, 2021.

\bibitem{shi2020multi}
M.~Shi, Y.~Tang, X.~Zhu, D.~Wilson, and J.~Liu, ``Multi-class imbalanced graph
  convolutional network learning,'' in \emph{Proceedings of the Twenty-Ninth
  International Joint Conference on Artificial Intelligence (IJCAI-20)}, 2020.

\bibitem{ravi2022wangiri}
A.~Ravi, M.~Msahli, H.~Qiu, G.~Memmi, A.~Bifet, and M.~Qiu, ``Wangiri fraud:
  Pattern analysis and machine learning-based detection,'' \emph{IEEE Internet
  of Things Journal}, 2022.

\bibitem{jiang2022telecom}
Y.~Jiang, G.~Liu, J.~Wu, and H.~Lin, ``Telecom fraud detection via
  hawkes-enhanced sequence model,'' \emph{IEEE Transactions on Knowledge and
  Data Engineering}, 2022.

\bibitem{hu2022btg}
X.~Hu, H.~Chen, S.~Liu, H.~Jiang, G.~Chu, and R.~Li, ``Btg: A bridge to graph
  machine learning in telecommunications fraud detection,'' \emph{Future
  Generation Computer Systems}, vol. 137, pp. 274--287, 2022.

\bibitem{krasic2022telecom}
I.~Krasi{\'c} and S.~{\v{C}}elar, ``Telecom fraud detection with machine
  learning on imbalanced dataset,'' in \emph{2022 International Conference on
  Software, Telecommunications and Computer Networks (SoftCOM)}.\hskip 1em plus
  0.5em minus 0.4em\relax IEEE, 2022, pp. 1--6.

\bibitem{2021GraphMix}
V.~Verma, M.~Qu, K.~Kawaguchi, A.~Lamb, Y.~Bengio, J.~Kannala, and J.~Tang,
  ``Graphmix: Improved training of gnns for semi-supervised learning,'' in
  \emph{National Conference on Artificial Intelligence}, 2021.

\bibitem{li2018adaptive}
R.~Li, S.~Wang, F.~Zhu, and J.~Huang, ``Adaptive graph convolutional neural
  networks,'' in \emph{Proceedings of the AAAI conference on artificial
  intelligence}, vol.~32, no.~1, 2018.

\bibitem{chen2019deep}
Y.~Chen, L.~Wu, and M.~J. Zaki, ``Deep iterative and adaptive learning for
  graph neural networks,'' \emph{arXiv preprint arXiv:1912.07832}, 2019.

\bibitem{liu2019neural}
W.~Liu, Z.~Liu, J.~M. Rehg, and L.~Song, ``Neural similarity learning,''
  \emph{Advances in Neural Information Processing Systems}, vol.~32, 2019.

\bibitem{dou2020enhancing}
Y.~Dou, Z.~Liu, L.~Sun, Y.~Deng, H.~Peng, and P.~S. Yu, ``Enhancing graph
  neural network-based fraud detectors against camouflaged fraudsters,'' in
  \emph{Proceedings of the 29th ACM International Conference on Information \&
  Knowledge Management}, 2020, pp. 315--324.

\bibitem{liu2019geniepath}
Z.~Liu, C.~Chen, L.~Li, J.~Zhou, X.~Li, L.~Song, and Y.~Qi, ``Geniepath: Graph
  neural networks with adaptive receptive paths,'' in \emph{Proceedings of the
  AAAI Conference on Artificial Intelligence}, vol.~33, no.~01, 2019, pp.
  4424--4431.

\bibitem{wang2019semi}
D.~Wang, J.~Lin, P.~Cui, Q.~Jia, Z.~Wang, Y.~Fang, Q.~Yu, J.~Zhou, S.~Yang, and
  Y.~Qi, ``A semi-supervised graph attentive network for financial fraud
  detection,'' in \emph{2019 IEEE International Conference on Data Mining
  (ICDM)}.\hskip 1em plus 0.5em minus 0.4em\relax IEEE, 2019, pp. 598--607.

\bibitem{liu2018heterogeneous}
Z.~Liu, C.~Chen, X.~Yang, J.~Zhou, X.~Li, and L.~Song, ``Heterogeneous graph
  neural networks for malicious account detection,'' in \emph{Proceedings of
  the 27th ACM International Conference on Information and Knowledge
  Management}, 2018, pp. 2077--2085.

\bibitem{hamilton2017inductive}
W.~Hamilton, Z.~Ying, and J.~Leskovec, ``Inductive representation learning on
  large graphs,'' \emph{Advances in neural information processing systems},
  vol.~30, 2017.

\bibitem{bartlett2006convexity}
P.~L. Bartlett, M.~I. Jordan, and J.~D. McAuliffe, ``Convexity, classification,
  and risk bounds,'' \emph{Journal of the American Statistical Association},
  vol. 101, no. 473, pp. 138--156, 2006.

\bibitem{beijbom2014guess}
O.~Beijbom, M.~Saberian, D.~Kriegman, and N.~Vasconcelos, ``Guess-averse loss
  functions for cost-sensitive multiclass boosting,'' in \emph{International
  Conference on Machine Learning}.\hskip 1em plus 0.5em minus 0.4em\relax PMLR,
  2014, pp. 586--594.

\bibitem{liu2019agrm}
M.~Liu, J.~Liao, J.~Wang, and Q.~Qi, ``Agrm: attention-based graph
  representation model for telecom fraud detection,'' in \emph{ICC 2019-2019
  IEEE International Conference on Communications (ICC)}.\hskip 1em plus 0.5em
  minus 0.4em\relax IEEE, 2019, pp. 1--6.

\bibitem{kipf2016semi}
T.~N. Kipf and M.~Welling, ``Semi-supervised classification with graph
  convolutional networks,'' \emph{arXiv preprint arXiv:1609.02907}, 2016.

\bibitem{velivckovic2017graph}
P.~Veli{\v{c}}kovi{\'c}, G.~Cucurull, A.~Casanova, A.~Romero, P.~Lio, and
  Y.~Bengio, ``Graph attention networks,'' \emph{arXiv preprint
  arXiv:1710.10903}, 2017.

\bibitem{wang2019fdgars}
J.~Wang, R.~Wen, C.~Wu, Y.~Huang, and J.~Xiong, ``Fdgars: Fraudster detection
  via graph convolutional networks in online app review system,'' in
  \emph{Companion proceedings of the 2019 World Wide Web conference}, 2019, pp.
  310--316.

\bibitem{zhang2019key}
Y.~Zhang, Y.~Fan, Y.~Ye, L.~Zhao, and C.~Shi, ``Key player identification in
  underground forums over attributed heterogeneous information network
  embedding framework,'' in \emph{Proceedings of the 28th ACM international
  conference on information and knowledge management}, 2019, pp. 549--558.

\bibitem{liu2020alleviating}
Z.~Liu, Y.~Dou, P.~S. Yu, Y.~Deng, and H.~Peng, ``Alleviating the inconsistency
  problem of applying graph neural network to fraud detection,'' in
  \emph{Proceedings of the 43rd international ACM SIGIR conference on research
  and development in information retrieval}, 2020, pp. 1569--1572.

\bibitem{hu2022gatcobo}
X.~Hu, H.~Chen, J.~Zhang, H.~Chen, S.~Liu, X.~Li, Y.~Wang, and X.~Xue,
  ``Gat-cobo: Cost-sensitive graph neural network for telecom fraud
  detection,'' \emph{arXiv preprint arXiv:2212.XXXXX}, 2022.

\end{thebibliography}

\begin{IEEEbiography}[{\includegraphics[width=1in,height=1.25in,clip,keepaspectratio]{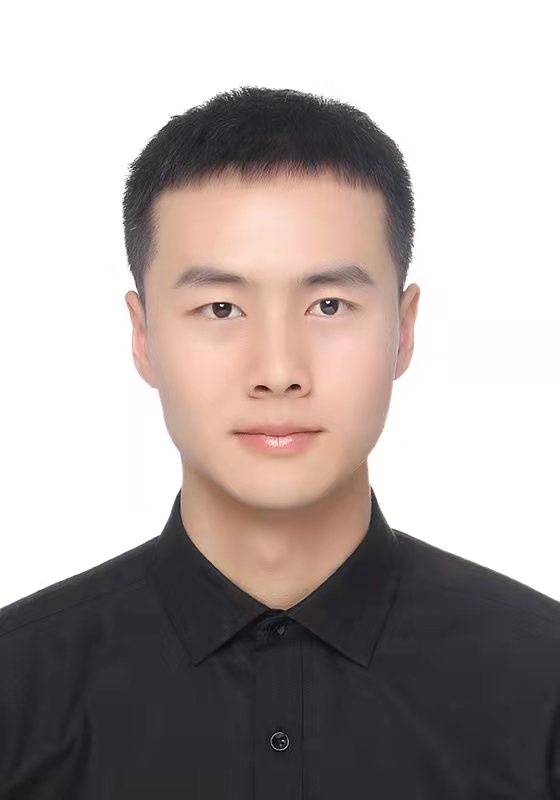}}]{Xinxin Hu}
 is currently a Doctor candidate at National Digital Switching System Engineering and Technological Research Center(NDSC) at Zhengzhou, China. He received the B.E. degree in Huazhong University of Science and Technology in 2017. His research interests include big data analysis, complex network, machine learning, mobile communication network security. 
\end{IEEEbiography}

\begin{IEEEbiography}[{\includegraphics[width=1in,height=1.25in,clip,keepaspectratio]{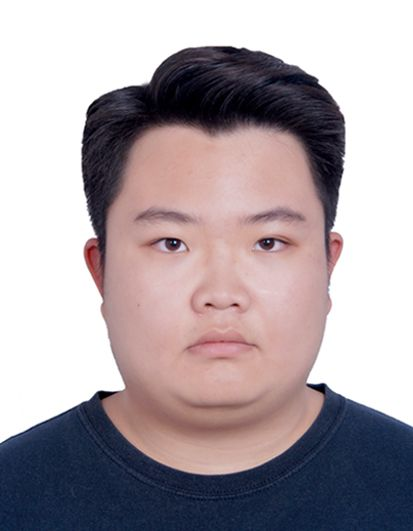}}]{Haotian Chen}
is currently a MEng student at The Edward S. Rogers Sr. Department of Electrical \& Computer Engineering, University of Toronto at Toronto, Canada. He received his Bachelor of computer science degree from University of Calgary. His research interests include machine learning, big data analysis, and network security.
\end{IEEEbiography}


\begin{IEEEbiography}[{\includegraphics[width=1in,height=1.25in,clip,keepaspectratio]{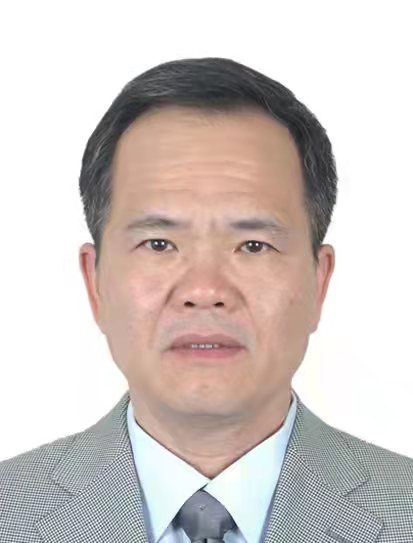}}]{Hongchang Chen}
is currently a professor at National Digital Switching System Engineering and Technological Research Center(NDSC). His research interests include future network communication and future network architecture.
\end{IEEEbiography}

\begin{IEEEbiography}[{\includegraphics[width=1in,height=1.25in,clip,keepaspectratio]{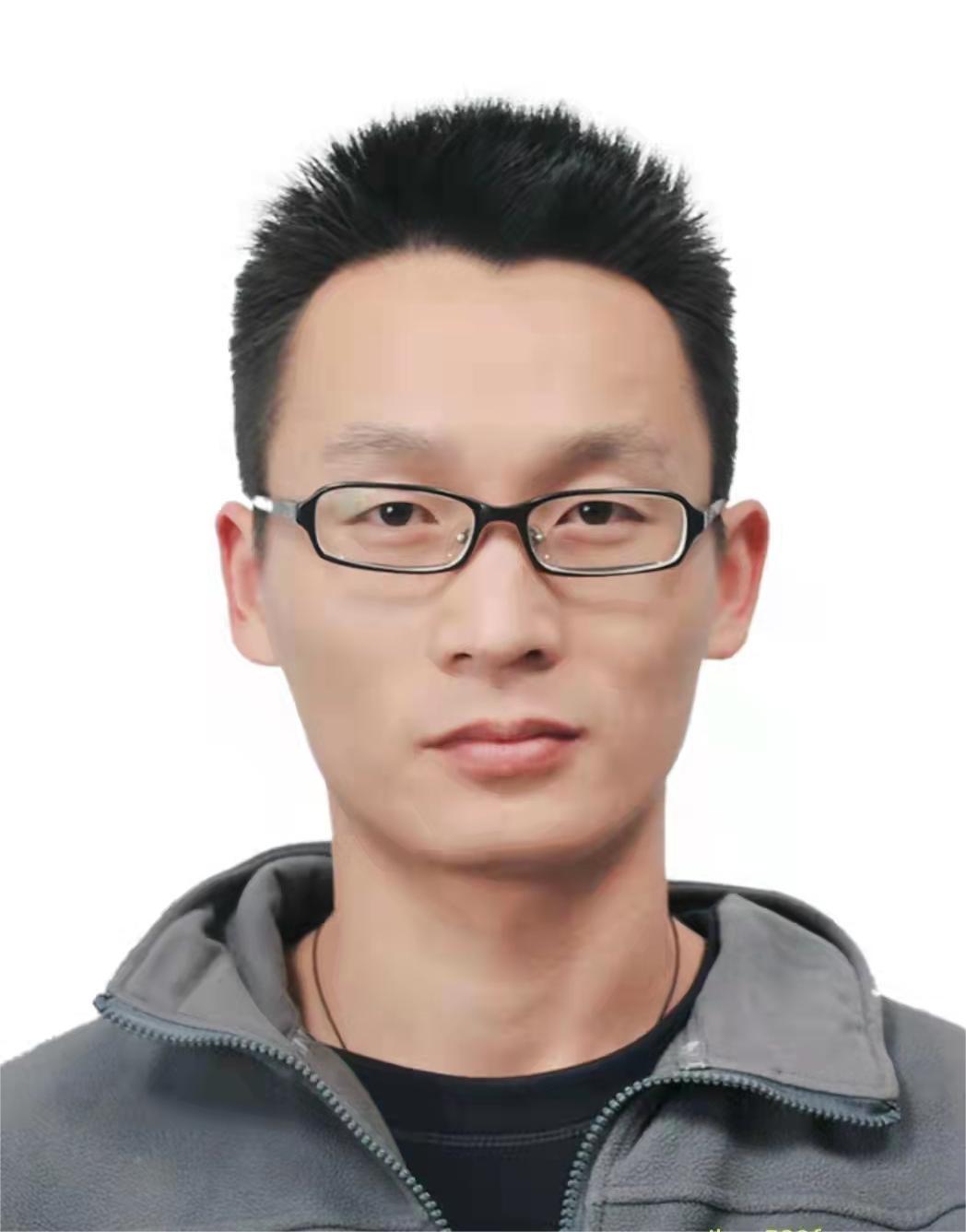}}]{Shuxin Liu}
is currently an Assistant Research Fellow with NDSC and the Director of the Laboratory of Network Architecture and Signaling Protocol Analysis. His research interests include network evolution, link prediction, network behavior analysis, and communication network security. 
\end{IEEEbiography}

\begin{IEEEbiography}[{\includegraphics[width=1in,height=1.25in,clip,keepaspectratio]{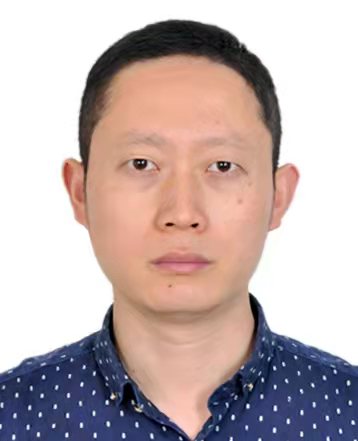}}]{Xing Li}
is  currently an assistant researcher of National Digital Switching System Engineering and Technological Research Center(NDSC), China. He received the Ph.D. degree from Information Engineering University in 2020. His major research interests include network science and cyberspace security.
\end{IEEEbiography}

\begin{IEEEbiography}[{\includegraphics[width=1in,height=1.25in,clip,keepaspectratio]{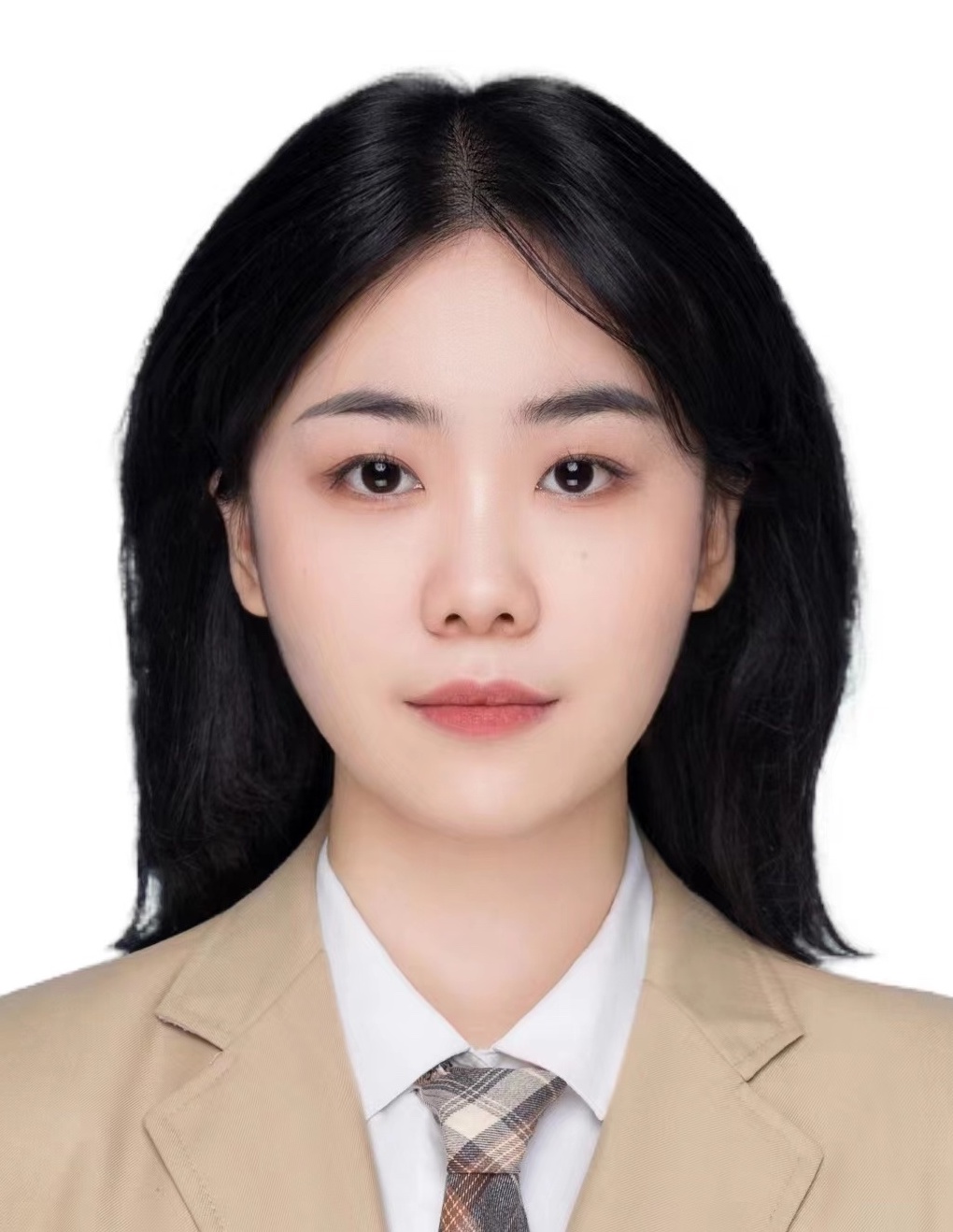}}]{Shibo Zhang}
is currently studying for a Ph.D. in National Digital Switching System Engineering and Technological Research Center(NDSC), China. She received the degree of Master of Science (Signal Processing) in Nanyang Technological University in 2021. Her research interests include complex network, machine learning.
\end{IEEEbiography}

\begin{IEEEbiography}[{\includegraphics[width=1in,height=1.25in,clip,keepaspectratio]{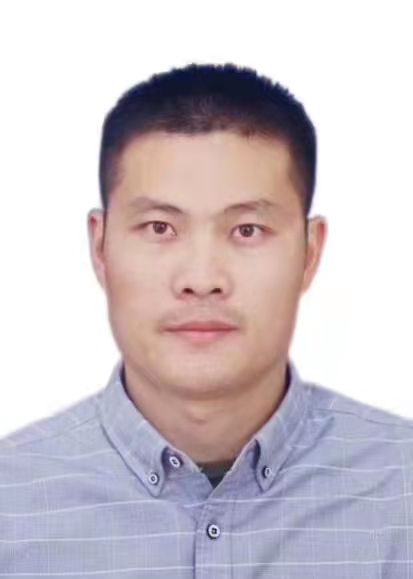}}]{Yahui Wang}
is currently a Doctor candidate at National Digital Switching System Engineering and Technological Research Center(NDSC) at Zhengzhou, China. He received the B.E. degree in Xi'an Jiaotong University in 2009. His research interests include data mining, complex network, machine learning.
\end{IEEEbiography}

\begin{IEEEbiography}[{\includegraphics[width=1in,height=1.25in,clip,keepaspectratio]{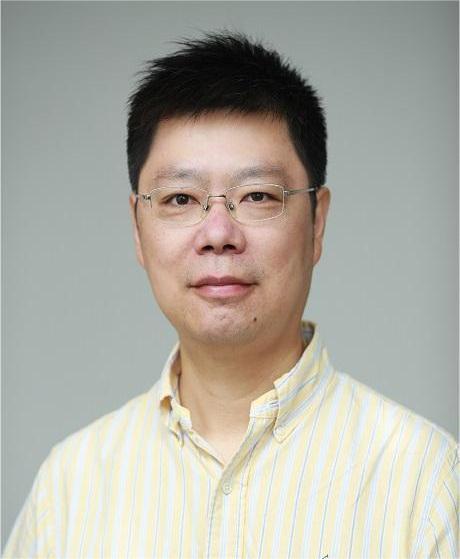}}]{Xiangyang Xue}
is currently a professor of computer science with Fudan University, Shanghai, China. He received the BS, MS, and PhD degrees in communication engineering from Xidian University, Xi'an, China, in 1989, 1992, and 1995, respectively. His research interests include multimedia information processing, big data analysis and machine learning. 
\end{IEEEbiography}

\end{document}